\documentclass[iop,twocolappendix]{emulateapj}
\usepackage[para]{threeparttable}
\usepackage{graphics}
\usepackage{color}

\usepackage{natbib}
\usepackage{float}

\begin{document}

\title{Line Identifications of Type I Supernovae: \\ On the Detection of \ion{Si}{2} for these Hydrogen-poor Events}

\author{J.~T.~Parrent\altaffilmark{1}, D.~Milisavljevic\altaffilmark{1}, A.~M.~Soderberg\altaffilmark{1}, M.~Parthasarathy\altaffilmark{2}}

\altaffiltext{1}{Harvard-Smithsonian Center for Astrophysics, 60 Garden St., Cambridge, MA 02138, USA}
\altaffiltext{2}{Indian Institute of Astrophysics, Koramangala, Bangalore, 560034, India}

\begin{abstract}

Here we revisit line identifications of type~I supernovae and highlight trace amounts of unburned hydrogen as an important free parameter for the composition of the progenitor. Most 1-dimensional stripped-envelope models of supernovae indicate that observed features near 6000~$-$~6400~\AA\ in type~I spectra are due to more than \ion{Si}{2}~$\lambda$6355. However, while an interpretation of conspicuous \ion{Si}{2}~$\lambda$6355 can approximate 6150~\AA\ absorption features for all type~Ia supernovae during the first month of free expansion, similar identifications applied to 6250~\AA\ features of type~Ib and Ic supernovae have not been as successful. When the corresponding synthetic spectra are compared to high quality time-series observations, the computed spectra are frequently too blue in wavelength. Some improvement can be achieved with \ion{Fe}{2} lines that contribute red-ward of 6150~\AA, however the computed spectra either remain too blue, or the spectrum only reaches fair agreement when the rise-time to peak brightness of the model conflicts with observations by a factor of two. This degree of disagreement brings into question the proposed explosion scenario. 
Similarly, a detection of strong \ion{Si}{2}~$\lambda$6355 in the spectra of broad-lined Ic and super-luminous events of type I/R is less convincing despite numerous model spectra used to show otherwise. Alternatively, we suggest 6000~$-$~6400~\AA\ features are possibly influenced by either trace amounts of hydrogen, or blue-shifted absorption and emission in H$\alpha$, the latter being an effect which is frequently observed in the spectra of hydrogen-rich, type~II supernovae. 

\end{abstract}


\section{Introduction}

Studying the evolution of a supernova's spectral energy distribution is a formidable task on account of excessive line blending that, for all intents and purposes, erases most traces of a reference continuum level. Specific line transitions that contribute to conspicuous features in SN~Ia spectra have been mostly identified, whereas the identification of weak spectral features for most SN~I supernovae remain uncertain (see Table~1). Line blending on the scale generated by the large expansion velocities ($>$~10$^{4}$~km~s$^{-1}$) thus creates a number of hurdles for supernova spectroscopists who are interested in extracting information about the composition of the ejecta (cf. \citealt{Payne36,Popper37}).  

In terms of a taxonomy, or a observed spectral sequence, supernovae of various luminosity classes are divided into two central categories, types~I~and~II, depending on whether or not the spectrum contains conspicuous signatures of hydrogen \citep{Minkowski41}. For type~II supernovae (SN~II), signatures of hydrogen Balmer lines come in the form of distinct P Cygni profiles. Depending on the decline in the light curve, type~II supernovae can be further sub-classified as SN~IIL for a linear decline and SN~IIP for sustained brightness before an eventual decline. However, drawing distinctions between SN~IIL~and~IIP is less obvious considering an existing continuum of observed properties \citep{Sanders14,Faran14,Pejcha15,Gall15,Chakraborti15}. 

Broadly speaking, hydrogen-poor SN~I supernovae can be separated into three subclasses, SN~Ia, Ib, and Ic. In addition, the  spectra of SN~IIb evolve from SN~II in appearance, with weak H$\alpha$, but later transition to a spectrum with lines of helium similar to SN~Ib (\citealt{Filippenko88}). Due to similarities between some caught-early SN~Ib to SN~IIb at later times, some SN~Ib may be caught-late SN~IIb (\citealt{Folatelli14}; see also \citealt{YLiu15}).

Types Ib and Ic are generally distinguished by the presence or absence of obvious helium signatures \citep{Wheeler95,Dessart15}. Additionally, thermonuclear SN~Ia reveal conspicuous signatures of \ion{Si}{2}~$\lambda$6355, while the latter varieties of SN~I core-collapse supernovae do not \citep{Filippenko97}; i.e. stripped-envelope subclasses of SN~Ib/Ic are defined by \citet{Filippenko97} and previous authors as not having a conspicuous doublet signature of \ion{Si}{2}~$\lambda$6355, which is a line detected for all SN~Ia \citep{Wheeler95}.

\begin{figure*}
\centerline{\includegraphics*[scale=0.47, trim = 0mm 0mm 0mm 0mm]{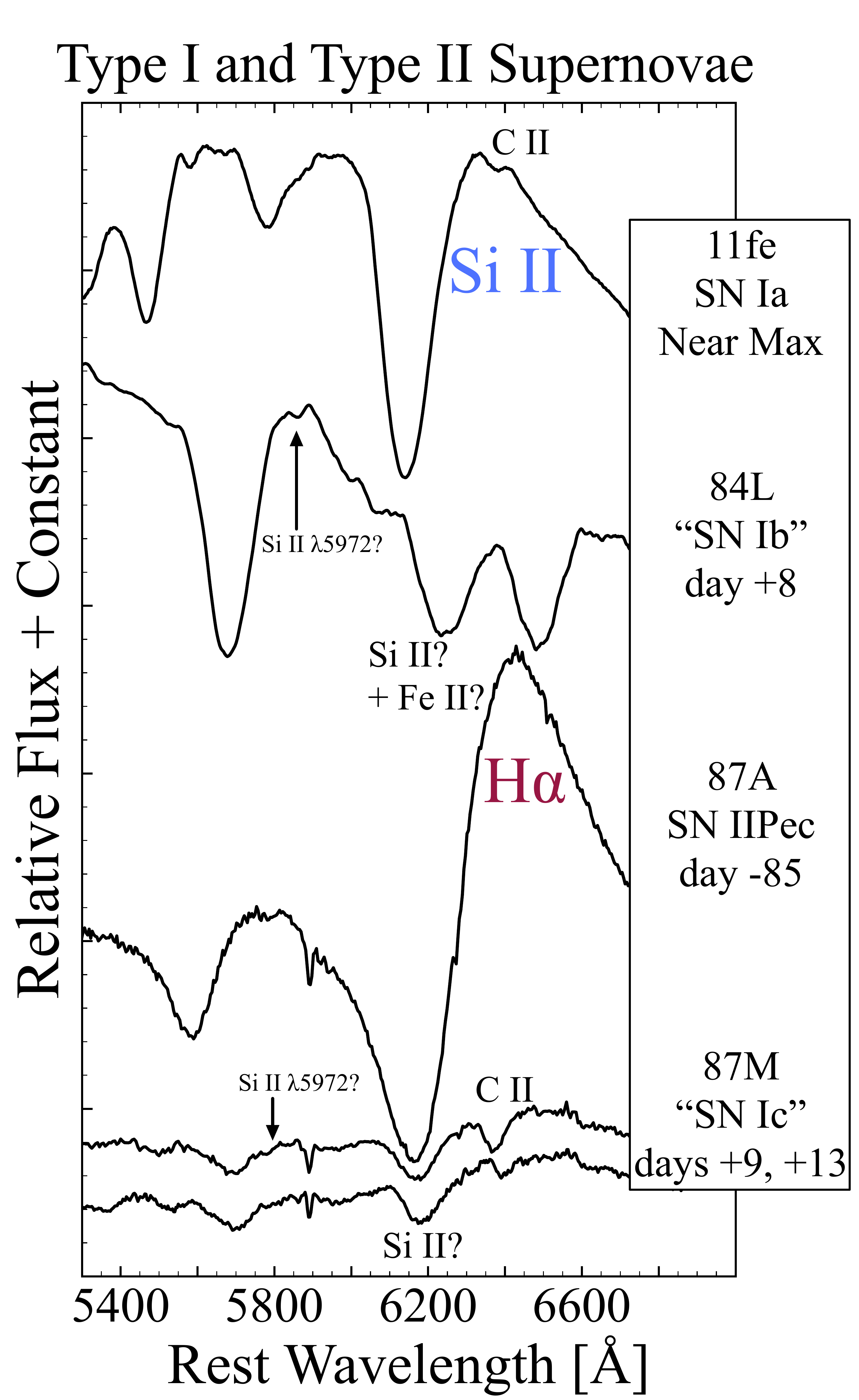}}
\caption{Comparisons of spectra from SN~1987A soon after the explosion on UT February 24, 1987 \citep{Pun95}; SN~1984L at approximately day~$+$8 relative to {\it B}-band maximum light \citep{Wheeler85,Doggett85,Tsvetkov87}; SN~1987M on UT September 28, 1987 and UT October 2, 1987, approximately days~$+$9 and $+$13, respectively, relative to {\it B}-band maximum light \citep{Filippenko90,Jeffery87M91}; and SN~2011fe near maximum light \citep{Pereira13}. Labels for SN~1987A and 2011fe denote ions for which detections have been verified, while labels for the 6250~\AA\ features of SN~1984L and 1987M represent unproven candidate ions. For SN~2011fe, 1984L, 1987A a single epoch is shown, whereas for SN 1987M
two epochs are shown ($+$9, $+$13).}
\label{Fig:Filippenko97_redo}
\end{figure*}

For a comparison of these types of supernovae, in Figure~\ref{Fig:Filippenko97_redo} we plot the SN~Ib~1984L, SN~IIpec~1987A, SN~Ic~1987M, and SN~Ia~2011fe and note the similarity of 6250~\AA\ features between SN~II-peculiar~1987A and SN~Ic~1987M is striking.  However, in the similar Fig.~1 of \citet{Filippenko97}, weak absorption features near 6250~\AA\ are labeled as \ion{Si}{2} for both ``SN~Ib''~1984L and ``SN~Ic''~1987M, implying the spectra harbor a lone and weak feature of silicon when this feature may instead be influenced by trace amounts of hydrogen for some SN~Ib \citep{James10}.

A precise definition of SN~Ib/Ic has since varied over the years (see \citealt{Fryer04,Gray09}). Some interpretations assume hydrogen is absent for SN~I ``by definition of being type~I'' (cf. \citealt{Maguire10,Eldridge13}), while lone detections of \ion{Si}{2}~$\lambda$6355 are deemed robust for SN~Ib/Ic, and therefore accurate for measurement (see, e.g., \citealt{Taubenberger06,Modjaz14}). Line identifications for 6000$-$6400~\AA\ features in the spectra of broad-lined SN~Ic (BL-Ic) and type~I/R super-luminous supernovae (SLSN) have also been debated. Most notably, estimates of \ion{Si}{2} and \ion{Fe}{2} conflict when 6000~$-$~6400~\AA\ spectral features are interpreted to be dominated by \ion{Si}{2}, while model spectra producing mostly signatures of \ion{Si}{2} are consistently too blue in wavelength when compared to observations (cf. \citealt{Mazzali97ef00,Inserra13,Lyman14}). 

The case for detectable hydrogen in the SN~Ic~1994I was discussed by \citet{Wheeler94}, while \citet{Clocchiatti96} claimed detections of weak lines of \ion{He}{1} as well. Both interpretations were later favored by \citet{Parrent07} who used \texttt{SYNOW} to show that a mix of both photospheric and higher velocity \ion{H}{1}~$\lambda$6563 offer better alternatives to the unmatched models of \citet{Branch06} and \citet{Sauer06}, both of which produce strong signatures of \ion{Si}{2}. The feature nearest to 6250~\AA\ has also been proposed to be a composite of \ion{C}{2}, \ion{Ne}{1}, and \ion{Si}{2} \citep{Hachinger12}. However, the abundance of \ion{Ne}{1} needed to match observations is reportedly too high to be physically consistent with the final abundances of candidate progenitor systems \citep{Ketchum08}.


\citet{Dessart15} recently produced non-LTE spectrum synthesis calculations, stemming from ``hydrogen-poor'' (M$_{H}$~$\sim$~10$^{-3}$~$-$~10$^{-2}$~M$_{\odot}$) and ``hydrogen-depleted'' models (M$_{H}$~$\sim$~10$^{-6}$~M$_{\odot}$), and have shown that a \ion{Si}{2}/\ion{Fe}{2}-weighted prescription for 6250~\AA\ features of SN~Ib/Ic may provide a more plausible alternative to treating weak signatures of H$\alpha$ as a free parameter. These models also indicate that if the progenitor star has above 10$^{-3}$~M$_{\odot}$ of hydrogen within a thin shell on its surface (i.e. not physically detached), then the corresponding post-explosion H$\alpha$ profile near 6250~\AA\ will resemble that of SN~IIb so long as the surface abundance of hydrogen is high enough. However, the rise-time of these SN Ib and SN Ic models are roughly twice as long as what is typically observed (cf. \citealt{Wheeler15}), which questions the models ability to represent ``standard compositions''. 

Finally, recent observations of the SN~Ib~2014C reveal evidence for interaction with a hydrogen-rich circumstellar medium (CSM) that was expelled in the years prior to explosion (\citealt{danmil15}; see also \citealt{Chugai06,Yan15,Moriya15}). Should we then expect trace amounts of hydrogen to remain within the outermost layers of the progenitor for some of these type~I core-collapse events?  

In this work we examine the latest model spectra for type~I supernovae and confront them with observations. In \S2 we highlight some current pitfalls when invoking detections and non-detections of select atomic species. In~\S3 we overview the ability of various explosion models to reproduce 6000$-$6400~\AA\ features in the spectra of SN~Ia, Ib, Ic, and more luminous events. 
In \S4 we summarize our findings.

\section{Spectral Line Identification}

Line blending is so severe for supernova spectra that it permits broad consistency between a synthesized spectrum and a majority of observed spectral features. These include absorption and emission line signatures that can be either conspicuous or weak below limiting noise. Some of these features can be approximated to a few lines, while others sustain a complex blend from first light to well after a year since explosion. 

Whether testing a benchmark model, or tracing observed features with a prescription of candidate lines, the interpretation of the observed spectral features, and the related empirical measurements, depend on the quality of the full fit from the supporting model. Moreover, if a model spectrum produces all but a single (or few) observed feature(s), the model is judged to be significantly incomplete. The same is true if the model produces spectral signatures that are not observed at the same epoch of the observation. Considering the relationships between the composition of the ejecta and the energy throughput, when a model's light curves are in good agreement with observations, the model spectra must also reflect observations before a model is sufficiently matched, and {\it vice versa}. 

Since boundaries between spectral line signatures are not well-defined for most supernovae, empirical false-positives are readily available within an underspecified parameter space of ``detections''.  Consequently, detections and non-detections of various ions and spectral lines do not follow classical definitions. For example, a non-detection of the strongest line from an ion at 5500~\AA\ does not necessarily imply a lack of significant spectral contamination from either this or other ions at 5500~\AA\ and other wavelengths. This can be made worse by low signal-to-noise ratio data and inadequate spectroscopic follow-up.

A common example is singly ionized silicon. This ion is considered detected in SN~Ia optical spectra when a number of spectral lines are conspicuous above noise and additional blending from adjacent lines. However, additional signatures of \ion{Si}{2}~$\lambda\lambda$3858, 5972 are not definitively detected for SN~Ib, Ic, and SLSN-I. 

\begin{figure*}
\centerline{\includegraphics[scale=0.6, trim = 0mm 0mm 110mm 0mm]{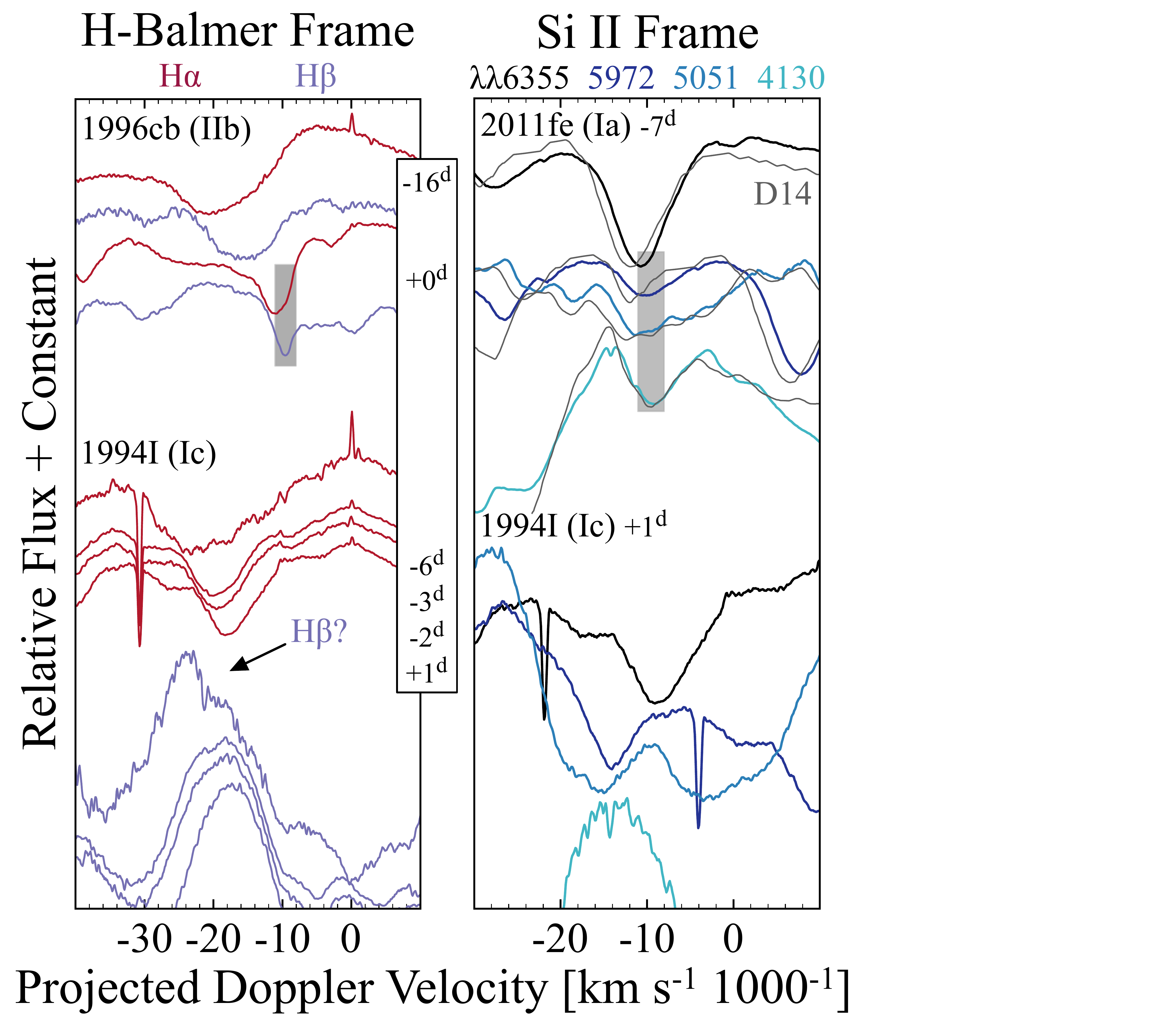}}
\caption{Plotted above are projected Doppler velocities $\mathcal{O}$(v/c) in the frame of the strongest H-Balmer lines (left) and \ion{Si}{2} lines (right). The light grey bands are meant to indicate the overlap between common signatures of a given ion. Spectrum references: SN~1994I \citep{Modjaz14}; SN~1996cb \citep{Matheson01}; SN~2011fe \citep{Pereira13}. Model references: (D14) PDDEL4, day $-$6.6 for SN~2011fe by \citet{Dessart14models}.}
\label{Fig:Hbeta}
\end{figure*}

See Figure~\ref{Fig:Hbeta} where we have plotted photospheric phase spectra of the SN~Ia~2011fe \citep{Pereira13} and the SN~Ic~1994I \citep{Filippenko95} in terms of \ion{Si}{2} and \ion{H}{1} line velocities $\mathcal{O}$(v/c). For SN~2011fe, its spectra are reasonably influenced by so-called conspicuous lines of \ion{Si}{2}. In comparison to SN~Ib/Ic, a lack of obvious \ion{Si}{2}~$\lambda$5972 is to be expected since the stronger \ion{Si}{2}~$\lambda$6355 would already be a weak signature compared to that for SN~Ia; i.e. a purported identification of \ion{Si}{2}~$\lambda$6355 so far has only one wavelength region by which to constrain the associated model.

Other ions with few strong lines, such as \ion{H}{1}, \ion{C}{2}, \ion{C}{3}, \ion{Na}{1}, and \ion{Si}{4}, face similar obstacles during photospheric phases as well, while a signature of \ion{Na}{1}~D is arguably absent for SN~Ia, minimal or absent for SN Ib, and significantly present for SN~Ic \citep{Dessart12,Dessart14,Dessart15}. Hence, studies of ejecta compositions benefit when the number of candidate lines, is either effectively maximized or close to the number of lines that are suspected to be present \citep{Baron96,Friesen14,Vacca15}. 

These are important points to consider when interpreting type~I spectra given that freshly synthesized ejecta are potentially contaminated by unburned and distinguishable traces of H, He, C, O, Si, Ca, and Fe-rich progenitor material near and above the primary volume of line formation (a so-called ``photosphere,'' see \citealt{HatanoAtlas,Branch05}). Additionally, an observed spectral signature associated with a particular ion at photospheric velocities can be influenced by an identical ion, in addition to other atomic species at similar or relatively higher velocities \citep{Marion13} (see Fig~3. of \citealt{Blondin15}).

For example, \citet{Sasdelli14} claim non-detections for all signatures of carbon in the early spectra of the peculiar SN~Ia~1991T without providing an explanation for a tentative detection of \ion{C}{3} $\lambda$4649. 
This identification is also consistent with kinetic energies deduced from detectable lines of \ion{Fe}{3} and a tentative detection of \ion{Si}{3}. For the computed spectrum at 4500~\AA, a feature is missing exactly where it is present for SN~1991T, 1997br, and 1999aa during early pre-maximum epochs. Therefore, abundance determinations of unburned carbon would benefit from knowing the subset of species giving rise to weak 4500~\AA\ features in the spectra of SN~1991T/1999aa-like events \citep{Hatano02,Parrent11,Parrent14}.

Detections of select atomic species are most apparent when line-blanketing effects are strong. A classic example is the absorption trough of \ion{Ti}{2} observed for SN~1991bg and 1999by on account of overall lower ejecta temperatures (\citealt{Doull11} and references therein). By contrast, less conspicuous detections of \ion{Ti}{2} will be blended and relatively ambiguous against a backdrop of signatures from other iron-peak elements. This is the root of the problem when sub-classifying supernovae according to their spectral features. 

As an example, \citet{White15} utilize non-detections of the \ion{Ti}{2} absorption trough observed blue-ward of 4450~\AA\ in an effort to distinguish peculiar SN~2002es-likes with conspicuous \ion{Ti}{2} \citep{Ganeshalignam12} from SN~2002cx-likes found without the classically seen trough of \ion{Ti}{2} absorption \citep{WLi03}. It is important to note, however, that despite claims of non-detections of \ion{Ti}{2}, the ejecta and spectra of SN~2002cx-likes are not likely void of \ion{Ti}{2} (see \citealt{Foley09,Kromer13,Stritzinger14}).

\begin{figure}
\centerline{\includegraphics*[scale=0.45, trim = 20mm 25mm 0mm 10mm]{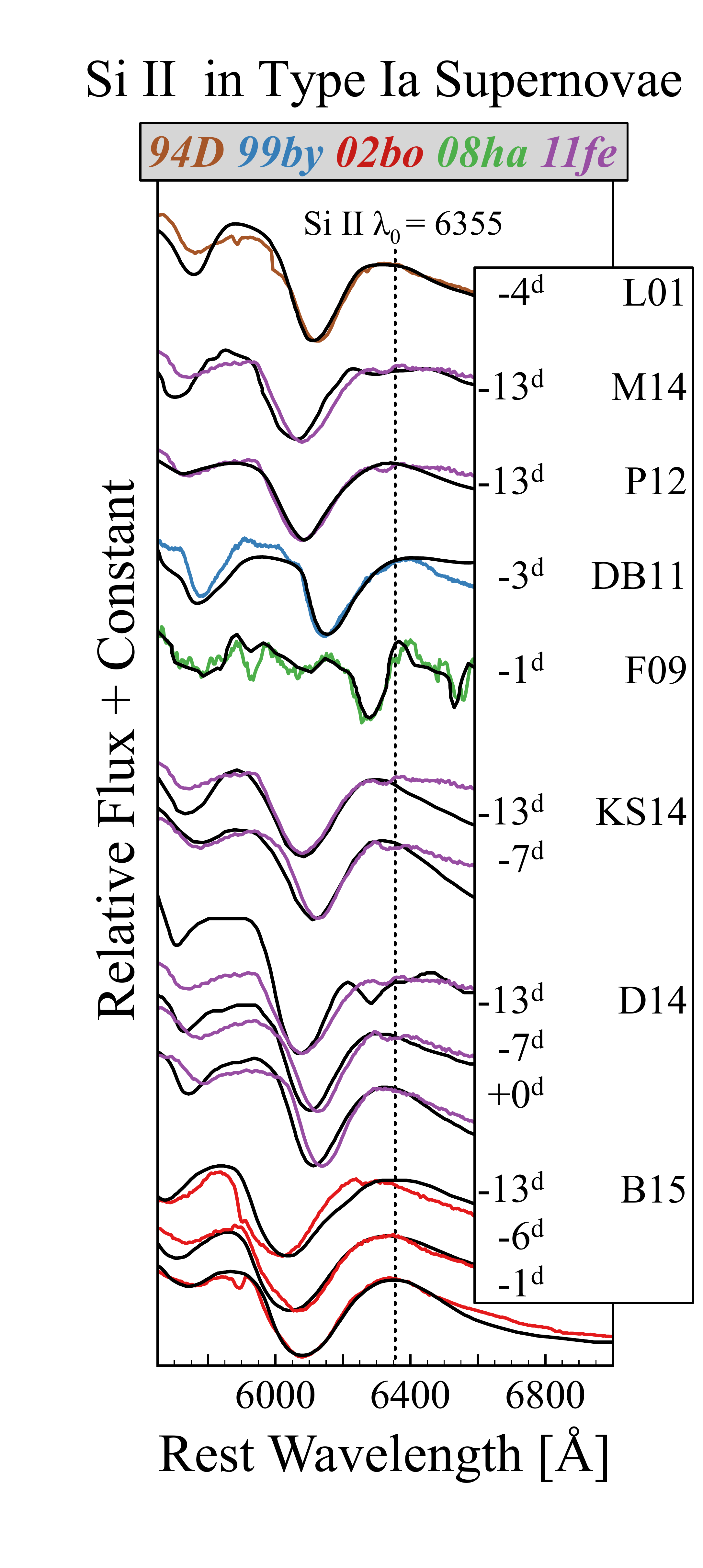}}
\caption{Comparisons between computed and observed spectra of select SN~Ia. Spectrum references: SN~1994D (brown), \citet{Patat96}; SN~1999by (blue), \citet{Garnavich04}; SN~2002bo (red), \citet{Benetti04}; SN~2008ha (green), \citet{Foley09}; SN~2011fe (purple), \citet{Parrent12,Pereira13,Mazzali14}. Model references: (L01) \citet{Lentz01a}; (M14) \citet{Mazzali14}; (P12) \citet{Parrent12}; (DB11) \citet{Doull11}; (F09) \citet{Foley09}; (KS14) \citet{Kerzendorf14}; (D14) \citet{Dessart14models}; (B15) \citet{Blondin15}.}
\label{Fig:SNIasilicon}
\end{figure}

\section{Model spectra of type~I supernovae \\near 6000$-$6400~\AA}

In this section we examine SN~I synthetic spectra found throughout the literature and assess respective goodness-of-fits when compared to observations. For each model spectrum shown, we have attempted to preserve the flux scaling for comparisons between observed and synthesized data in the original works, and we have normalized the spectra with respect to the 6000$-$6400~\AA\ profiles where appropriate. (This underscores the need for releasing all published model spectra and photometry.)  

\subsection{Thermonuclear SN Ia}

Plotted in Figure~\ref{Fig:SNIasilicon} are observations of SN~Ia subtypes compared to the computed spectra from various spectrum synthesizers at wavelengths spanning 5200~$-$~7000~\AA. All model spectra, generated from semi-empirical to full non-LTE radiation transport solvers, are capable of providing close match to the feature centered about 6150~\AA\ through prescriptions based on \ion{Si}{2}~$\lambda$6355. 

Specifically, all absorption and emission components are well-centered with the data, while the curvature of the profile wings are relatively well-matched. This is also in spite of either a ``non-standard'' distribution of \ion{Si}{2}, or two distinct components of photospheric and detached \ion{Si}{2}~$\lambda$6355. That is, precise and accurate predictions are achievable for the red and blue wings, which are both sensitive to mean expansion velocities and the radial decline in density.

To further appreciate the limiting accuracy of 6150~\AA\ features in SN~Ia spectra as having been identified as \ion{Si}{2}~$\lambda$6355, consider that an intrinsic difference of 500, 1000, and 2000~km~s$^{-1}$ in terms of line velocities would be reflected by an underlying shift of $\sim$~10, 20, and 40~\AA, respectively. That is, apart from opacity effects, significant differences of 2000~km~s$^{-1}$ can give rise to 40~\AA\ blue-shifts that are of similar order to redshifts seen for 6150~\AA\ features during the first month of free expansion. Despite these differences, model spectra for SN~Ia consistently reflect a suite of observations.

When confronted by events such as the SN~Ia~2012fr \citep{Childress13,Maund13}, where the conspicuousness of a higher velocity signature of \ion{Si}{2}~$\lambda$6355 is sensitive to the cadence of follow-up observations, approximate consistency with composite features near 6150~\AA\ is relatively large for SN~Ia. In other words, the relative location of high velocity \ion{Si}{2} would not be well-constrained without the accompanying time-series spectra. As we will discuss in \S3.4, this window of consistency is even greater for broad-lined SN~Ic (BL-Ic) and SLSN.

One reason a model SN~Ia might not match near the emission component of the \ion{Si}{2} is when the model does not center \ion{C}{2}~$\lambda$6580 at similar projected Doppler velocities as \ion{Si}{2} during the first month after the explosion \citep{Taubenberger11,Pereira13}. Other lines that impact this region for SN~Ia include \ion{S}{2}~$\lambda$6305 and lines of \ion{Fe}{2} near maximum light and thereafter. Despite these additional biases on direct measures of \ion{Si}{2} features, the identification of \ion{Si}{2}~$\lambda$6355 as a dominant line in the spectra of SN~Ia is robust for a variety of normal and peculiar events.

\subsection{Core-collapse Supernovae}

Based on the recent spectrum synthesis calculations from \citet{Dessart15}, who utilize a modified version of \texttt{CMFGEN} \citep{Hillier12}, stripped-envelope models that produce weak signatures of \ion{Si}{2}~$\lambda$6355 may avoid both consistent mismatch and the need for trace hydrogen by relying on contribution from \ion{Fe}{2}, as well as \ion{He}{1} in some instances \citep{Dessart12}. In this case, \ion{Fe}{2} effectively moves a signature dominated by \ion{Si}{2} toward redder wavelengths, thereby making a blend of \ion{Si}{2} and \ion{Fe}{2} a strong contender for 6250~\AA\ features of SN~Ib/Ic spectra. 

\citet{Dessart15} also draw association with the observed spectral sequence of SN~IIb, Ib, and Ic, but without comparing any of these models to observations. Notably, the SN~Ib 6p5Ax1 model of \citet{Dessart15} has a rise-time of 40 days. This rise-time is nearly twice as long as those typically observed for SN~Ib, while the SN~Ic 5p11Ax1 from \citet{Dessart15} has a rise-time that conflicts with most observations as well (cf. \citealt{Wheeler15}). The ``Bmf5p09'' series of models presented by \citet{Dessart12} also exhibit rise-times that conflict with observations by as much as 30 days.

\begin{figure*}
\centerline{\includegraphics*[scale=0.27]{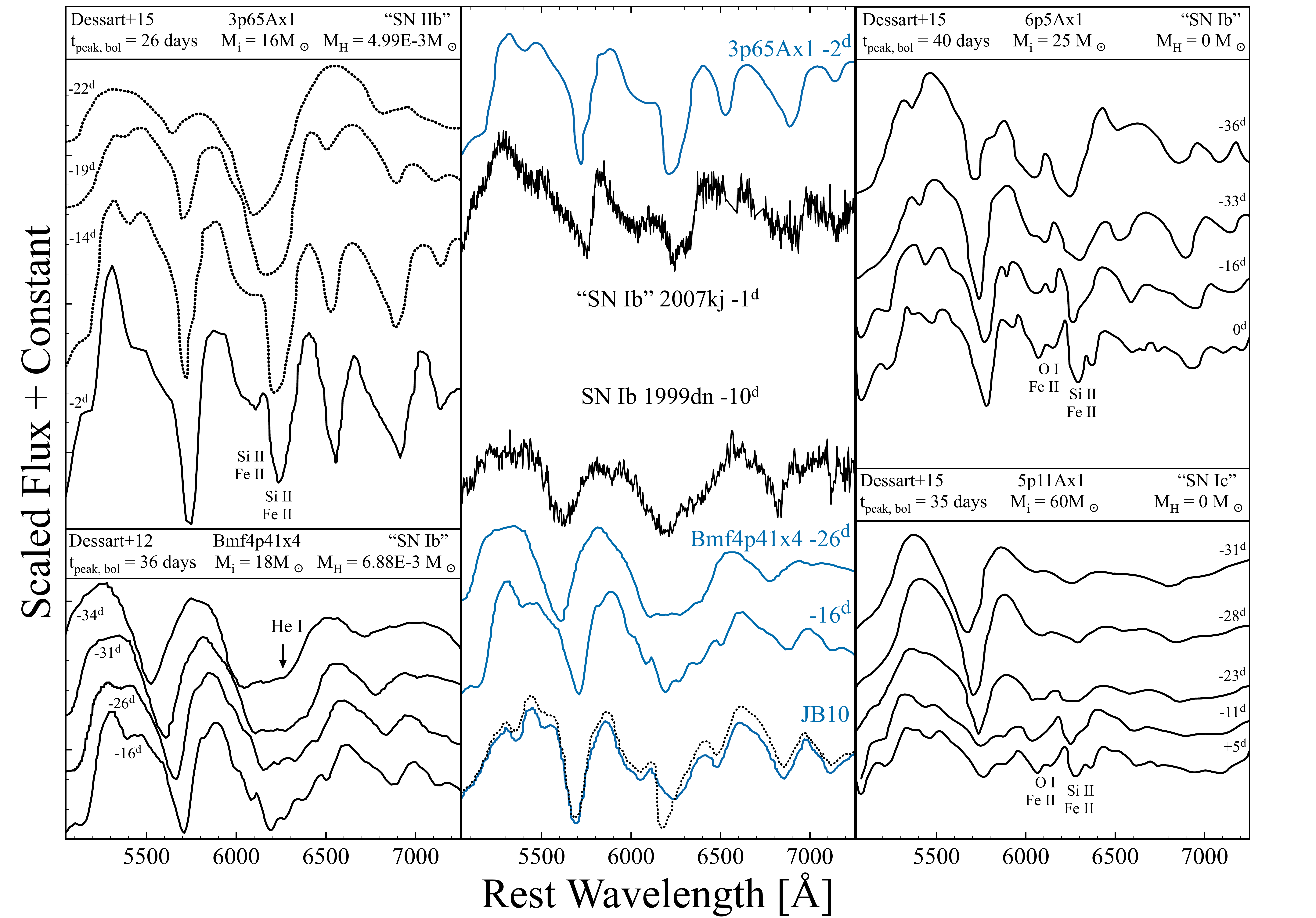}}
\caption{Comparisons between computed spectra, SN~2007kj \citep{Modjaz14}, and SN~1999dn \citep{Deng00}. The data have been scaled and normalized according to \citet{Jeffery07}. In the four corner panels, we plot time-series spectra of the SN IIb 3p65Ax1, SN~Ib 6p5Ax1, and SN~Ic 5p11Ax1 models presented by \citet{Dessart15}, and also include the earliest spectra of the ``standard SN~Ib'' Bmf4p41x4 presented by \citet{Dessart12}. The rise-time, initial mass, and final hydrogen mass for each model are listed in each panel. Dotted lines denote when the spectrum is contaminated by signatures of H$\alpha$. Certain features are labeled by ions producing significant structure within the computed spectrum. JB10 \texttt{PHOENIX} calculation of \citet{James10} with and without hydrogen are shown as the dashed black and solid blue line, respectively.}
\label{Fig:dessart1}
\end{figure*}

Since the modified version of \texttt{CMFGEN} used for supernovae has not been made publicly available, we are unable to test areas of improvement for any of the proposed models. The model spectra from \citet{Dessart12,Dessart15} have also not been made available for comparative studies, which was done for the SN~Ia models computed by \citet{Blondin15}. Therefore we take the liberty of digitizing model spectra of interest to plot them against observations in Figures~\ref{Fig:dessart1}~$-$~\ref{Fig:dessart3}. 



\subsubsection{``SN Ib'' 2007kj}

In an effort to distinguish SN~IIb, Ib, and Ic from their spectra, \citet{YLiu15} recently conducted a comparative study of ``the largest spectroscopic dataset'' of stripped-envelope supernovae. Using the Supernova Identification tool, \texttt{SNID} \citep{Blondin07}, and pseudo equivalent measurements of several spectral features, \citet{YLiu15} classify SN~2007kj as a SN~Ib. \citet{Modjaz14} previously used \texttt{SNID} to classify SN 2007kj as a SN~Ib as well.

However, comparison to the SN~IIb 3p65Ax1 model spectrum presented by \citet{Dessart15}, plotted in our Figure~\ref{Fig:dessart1}, would imply that SN 2007kj is instead either a caught-late or slow-evolving SN~IIb. That is, in spite of having incorporated the largest set of \texttt{SNID} templates to classify SN~IIb, Ib, and Ic, \citet{Modjaz14} and \citet{YLiu15} are unable to type SN~2007kj as a SN~IIb. 

Furthermore, if the SN~IIb 3p65Ax1 model had been built for SN~2007kj, the match to SN~2007kj would suggest the progenitor had $\sim$~10$^{-3}$~$-$~10$^{-2}$~M$_{\odot}$ of hydrogen on its surface at the time of explosion. It is also important to note that the 6250~\AA\ feature on day~$-$2 for the SN~IIb 3p65Ax1 model is not contaminated by H$\alpha$. This is in spite of a similar 6250~\AA\ feature on day~$-$14 (and earlier) where an H$\alpha$ signature is present for the model. By maximum light, this feature is predicted to be dominated by profiles of \ion{Si}{2} and \ion{Fe}{2} \citep{Baron95,Dessart15}.

\subsubsection{SN~Ib~1999dn}

The computed spectra of the SN Ib Bmf4p41x4 model are also associated with a progenitor having $\sim$~10$^{-3}$~M$_{\odot}$ of hydrogen near its surface \citep{Dessart12}. See the bottom-left panel of our Figures~\ref{Fig:dessart1}~and~\ref{Fig:dessart3}. The difference to note between the day~$-$2 spectrum of the aforementioned SN~IIb 3p65Ax1 and SN~Ib Bmf4p41x4 models is that the shape of the 6250~\AA\ feature in the latter is not SN~II-like, nor is its spectrum ``H$\alpha$-positive''. This is in spite of the hydrogen present in the ejecta, 
while the 6250~\AA\ feature is also blended with \ion{He}{1}. 

Looking at the middle panel of Figure~\ref{Fig:dessart1}, we find a promising match between the SN~Ib~1999dn at day~$-$10 to the SN~Ib Bmf4p41x4 model at day~$-$16. However, this model spectrum is not well-matched since the \ion{He}{1} features near 5650~\AA\ and 6900~\AA\ are significantly too red in wavelength. Increasing overall kinetic energy for the model, among other variables, can resolve this discrepancy. However, the issue that remains is whether or not this increase would result in a 6250~\AA\ feature that is too blue in wavelength. Even so, if a match is deemed secure for a model such as Bmf4p41x4, then this would imply the progenitor had $\sim$~10$^{-3}$~M$_{\odot}$ of remaining hydrogen in its outermost layers, as a SN Ib.

Alternatively, \citet{James10} previously presented \texttt{PHOENIX} calculations to show that unburned hydrogen at high expansion velocities (in a shell) is reasonable for SN~1999dn. (This comparison is also shown in Figure~\ref{Fig:dessart1}.) In this paradigm, the detachment velocity of the hydrogen shell is treated as a free parameter. With the shell displaced to 19,000~km~s$^{-1}$,
the material needed in the atmosphere of SN~Ib~1999dn to reproduce the 6250~\AA\ feature (M$_{H}$~$\lesssim$~10$^{-3}$~M$_{\odot}$) is less than 0.01~M$_{\odot}$$<$~M$_{H}$~$\lesssim$~0.2~M$_{\odot}$ inferred for SN~IIb~1993J by \citet{Baron95}. This mass estimate does not also conflict with the swept-up 5~x~10$^{-2}$~M$_{\odot}$ estimated from model comparisons to late-nebular spectra of SN~IIb \citep{Jerkstrand15}.  

\begin{figure*}
\centerline{\includegraphics*[scale=0.27]{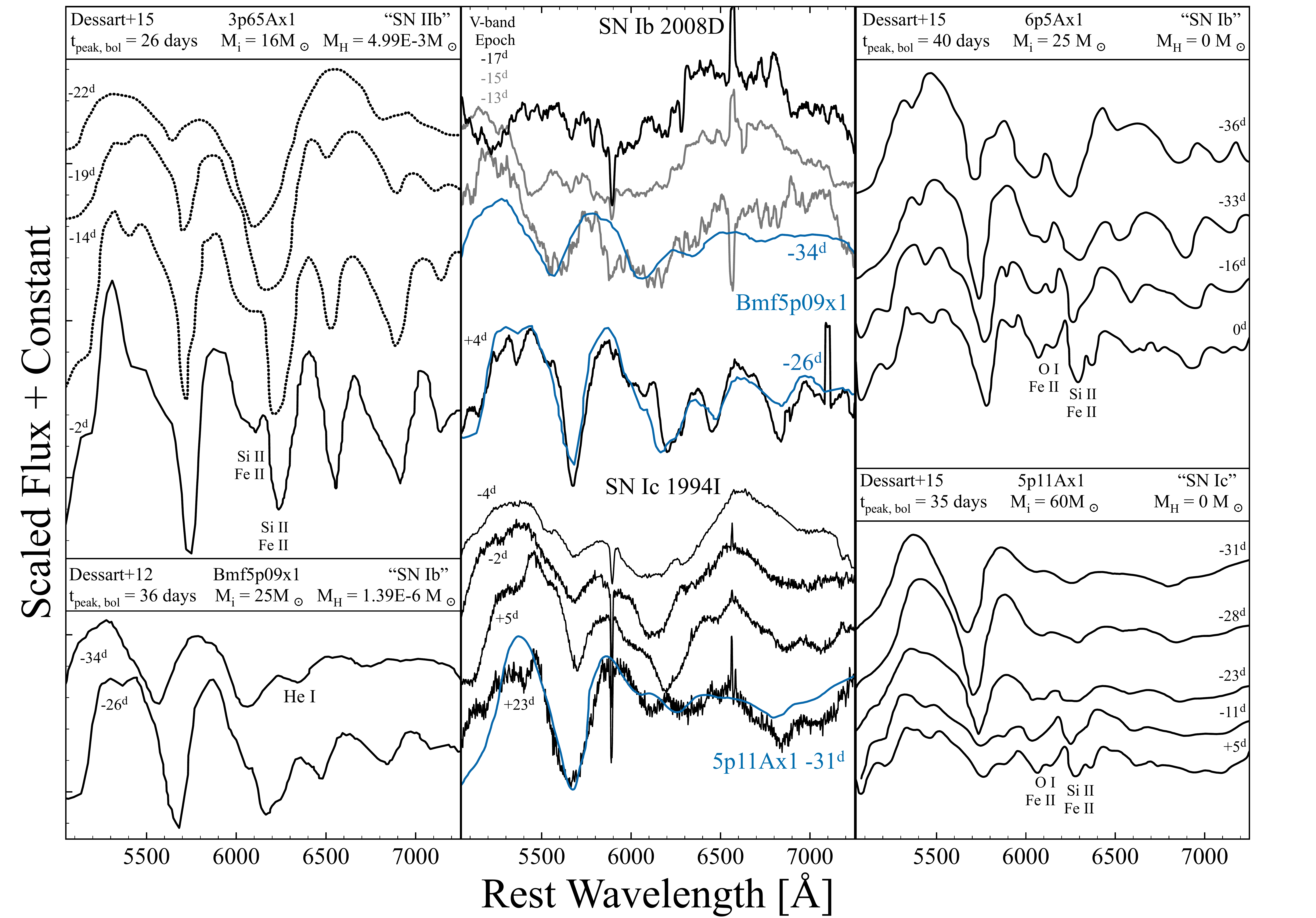}}
\caption{Same as Figure~\ref{Fig:dessart1} except now for SN~2008D \citep{Soderberg08D,Modjaz09} and SN~1994I \citep{Modjaz14}. We have replaced the Bmf4p41x4 spectra with those of Bmf5p09x1 from \citet{Dessart12}.}
\label{Fig:dessart2}
\end{figure*}

\subsubsection{SN Ib 2008D}

\citet{Dessart12} compared their Bmf5p09x4 model corresponding to 10.3 days since explosion to the February 2 spectrum of the SN~Ib~2008D. In terms of the number of days since maximum light, and with a rise-time of 36 days for the model, this corresponds to drawing a comparison between a model at day~$-$26 and an observed event at day~$+$4, a difference of one month. 

In Figure~\ref{Fig:dessart2} we compare the day $-$34 spectrum of Bmf5p09x4 (the earliest available) to the day $-$13, $-$15, and $-$17 observations of SN~2008D. Apart from the discordant epochs, the model spectrum does not confidently resemble those of SN~2008D prior to maximum light. The data are also fairly noisy at these early epochs, however this particular model is not optimized for drawing conclusions on the validity of \ion{Si}{2}~$\lambda$6355 for SN~2008D.

If both computed and observed spectra were those of SN~Ia, the differences are enough to signify different subtypes that may or may not be from the same family of progenitors. Specifically, and compared to SN~Ia where detections of \ion{Si}{2} have been verified, the earliest computed spectrum of Bmf5p09x4 does not overlap the 6250~\AA\ feature in the spectrum of SN~2008D near the date of explosion. One expects improvement through either stronger contribution from \ion{Fe}{2} or lower mean expansion velocities attributed to \ion{Si}{2}, or the model is flawed {\it ab initio}.

\subsubsection{``SN~Ic'' 1994I}

\begin{figure}
\centerline{\includegraphics[scale=0.3, trim = 20mm 0mm 70mm 25mm]{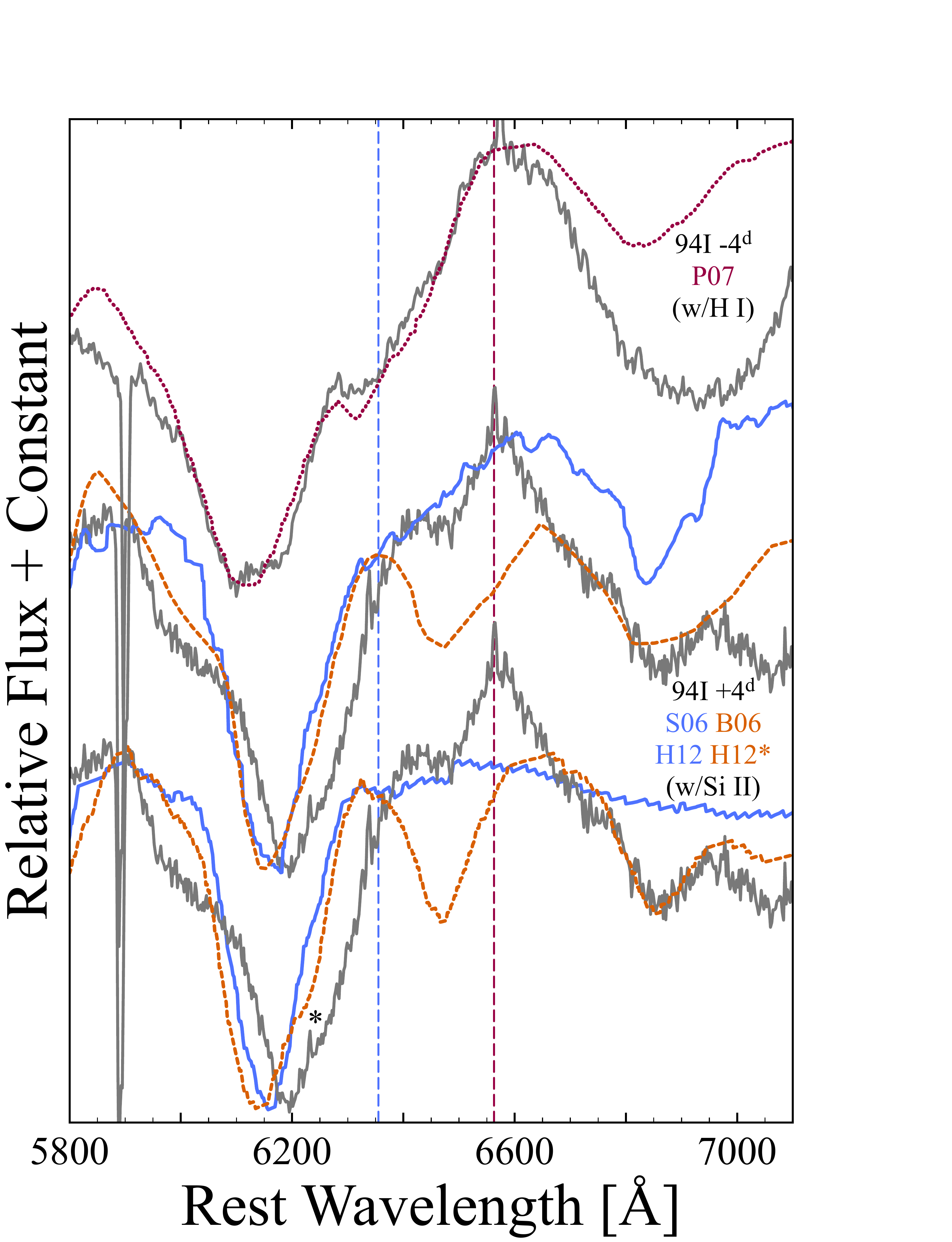}}
\caption{Model spectra compared to SN~Ic~1994I \citep{Modjaz14}. Model references: (P07) \texttt{SYNOW} calculation of \citet{Parrent07} with hydrogen treated as a free parameter; S06 \citep{Sauer06}; B06 \citep{Branch06}; H12 models are from \citet{Hachinger12}. The blue model, H12, is the base SN~Ic 94I model 16 days after explosion in the top of their Fig.~11 in green. The orange model, H12*, corresponds to the hydrogen-enriched SN~Ib in the top of their Fig.~11 in red. The black asterisk is near the location of H$\alpha$ coming through in this model. Like S06 and B06, both H12 models fail to match the minimum 6250~\AA\ in SN~Ic 1994I as they too are reliant on significant contribution from \ion{Si}{2}~$\lambda$6355. Vertical-dashed lines denote the rest wavelength of \ion{Si}{2}~$\lambda$6355~(blue) and \ion{H}{1}~$\lambda$6563~(red).}
\label{Fig:94I}
\end{figure}

For SN~1994I, a so-called ``standard SN~Ic'', two opposing interpretations prevail in the literature for two models that produce mismatched signatures of \ion{Si}{2} compared to observations (cf. \citealt{Branch06,Sauer06}). See Figure~\ref{Fig:94I} where we have over-plotted these two independent results.

\begin{figure*}
\centerline{\includegraphics*[scale=0.27]{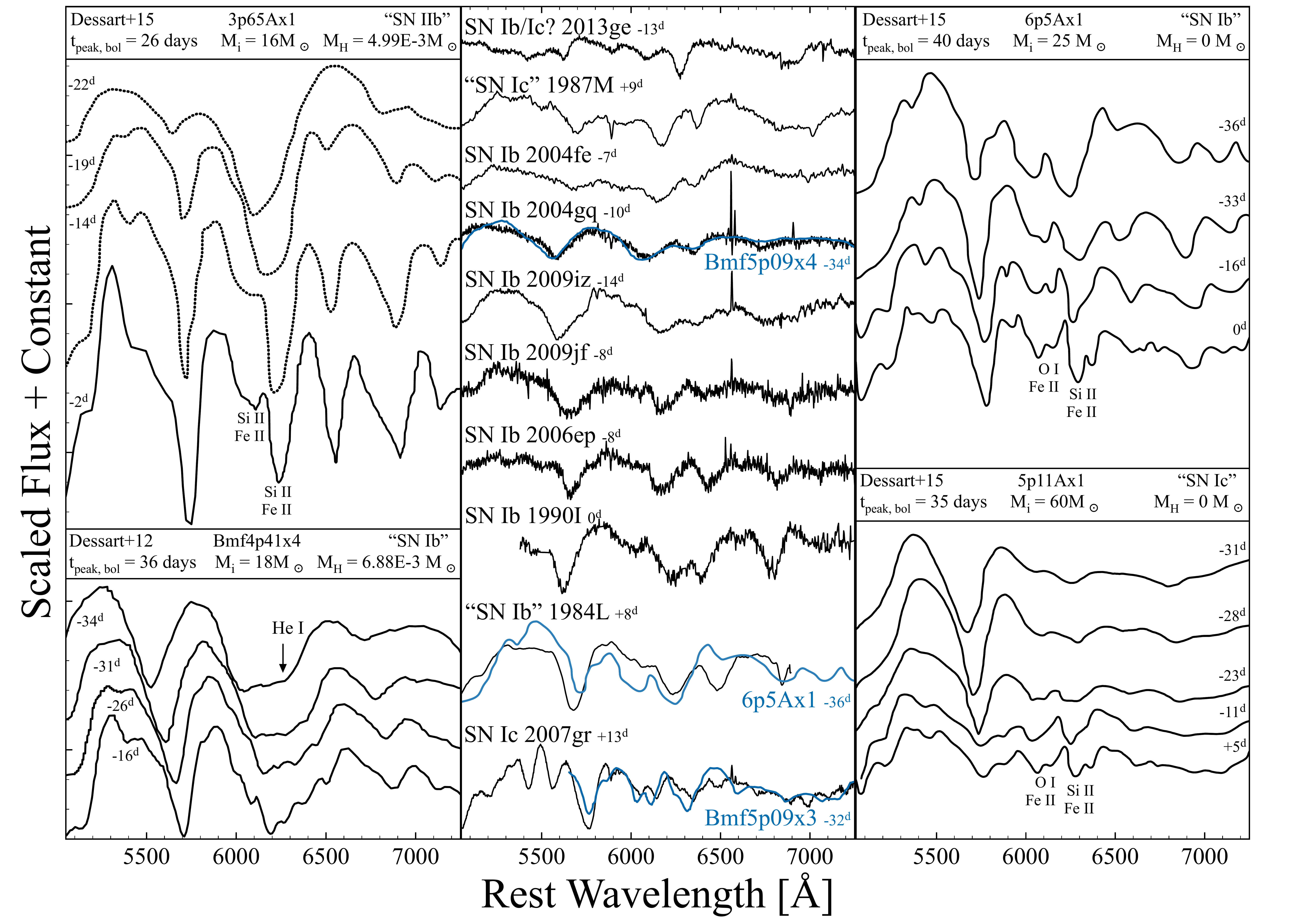}}
\caption{Same as Figure~\ref{Fig:dessart1} except now for select SN~Ib/Ic. Spectrum references: SN~1984L \citep{Wheeler85}; SN~1987M \citep{Filippenko90,Jeffery87M91}; SN~1990I \citep{Elmhamdi04}; 2004fe, 2004gq, SN~2006ge, SN~2007gr (telluric corrected), SN~2009iz, SN~2009jf (\citealt{Modjaz14}, see also \citealt{Valenti11}); SN~2013ge \citep{Drout15}.}
\label{Fig:dessart3}
\end{figure*}

\citet{Sauer06} predict the feature centered about 6200~\AA\ in the photospheric phase spectra of SN~1994I is primarily a product of \ion{C}{2}, \ion{Ne}{1}, and \ion{Si}{2}, with mostly \ion{Si}{2} present near maximum light. Based on their series of model spectra, including the fit near maximum light shown in Figure~\ref{Fig:94I}, \citet{Sauer06} claim a satisfactory match has been made. 

\citet{Branch06} previously used \texttt{SYNOW} to show that a satisfactory solution can be obtained for SN~1994I near maximum light when \ion{He}{1}, \ion{O}{1}, \ion{Ca}{2}, \ion{Ti}{2}, and \ion{Fe}{2} are assumed to form within similar strata of ejecta. When \citet{Branch06} incorporate \ion{Si}{2} into the \texttt{SYNOW} fit, the predicted offset of the feature dominated by \ion{Si}{2} is nearly identical to that later found by \citet{Sauer06}. Thus, while it is reasonable to expect a multi-component prescription throughout the spectra of SN~Ib/Ic, it is unreasonable to suspect a correct interpretation for over-prescribed \ion{C}{2}, \ion{Ne}{1}, and \ion{Si}{2} when the model used to support this prescription is incapable of producing the complex evolution of the data (cf. \citealt{Baron99,Ketchum08}).

\citet{Hachinger12} later used the model of \citet{Iwamoto94} to argue that \ion{Si}{2}~$\lambda$6355 is a viable option for the 6250~\AA\ feature in SN~1994I, and that this signature of \ion{Si}{2} can be contaminated by other species including H$\alpha$. However, despite offering to remedy the mismatch of model spectra with increased contribution from H$\alpha$, the hydrogen/helium-enriched C$+$O models of \citet{Hachinger12} remain dominated by ill-matched \ion{Si}{2}~$\lambda$6355. See Figure~\ref{Fig:94I} and note the asterisk next to the orange H12* spectrum near 6200~\AA; this dip is the H$\alpha$ coming through. 

In the top of Figure~\ref{Fig:94I}, we have plotted the fit produced by \citet{Parrent07} with \texttt{SYNOW} where contribution from H$\alpha$ was treated as a free parameter. It is important to reiterate that the assumptions of the \texttt{SYNOW/SYNAPPS} model are not realistic. Therefore best guesses for a composition, without any prior information from more detailed investigations, cannot necessarily be considered superior on account of a well-matched output spectrum. However, the primary purpose of \texttt{SYNOW}-like tools is to quickly explore where detailed models are consistently unsuccessful.


To compare with the more recent ``SN~Ic'' calculations of \citet{Dessart15}, in Figure~\ref{Fig:dessart2} we have included time-series observations of SN~1994I and over-plot the earliest computed spectrum of 5p11Ax1 at day~$-$31 with the day $+$23 spectrum of SN~1994I. Overall, one can see that the match is promising. However, the early evolution of the 6250~\AA\ feature observed for SN~1994I remains unaccounted for by the model. The 6250~\AA\ feature at a particular epoch may be dominated by \ion{Si}{2} and \ion{Fe}{2}, however there are no spectra of 5p11Ax1 prior to day~$-$31 to conclude whether this is the case.

\subsubsection{Other SN Ib/Ic}

Plotted in Figure~\ref{Fig:dessart3} are the spectra of several SN~Ib, SN~Ic, and SN~Ib/Ic-like events, e.g., SN~1984L, 1987M, and 2013ge \citep{Drout15}. Similar to Figures~\ref{Fig:dessart1}~and~\ref{Fig:dessart2}, we have included comparisons to select model spectra from \citet{Dessart12,Dessart15}.

In the middle panel of Figure~\ref{Fig:dessart3}, we see the day~$-$34 Bmf5p09x4 model spectrum from \citet{Dessart12} is most comparable to the SN~Ib~2004gq, yet the blue-ward offset in wavelength for the 6250~\AA\ feature is evident. Events that are similar to SN~2004gq include SN~Ib~2004fe, 2009iz, 2009jf, and 2006ep during pre-maximum epochs. For the SN~Ib~1990I, the feature nearest to 6250~\AA\ on its red-most side is likely blended with a signature of \ion{He}{1}~$\lambda$6678, and possibly with some contribution from \ion{C}{2}. 

For the spectrum of the SN~Ib~1984L, a \ion{He}{1}~$\lambda$5876 feature centered about $\sim$~5680~\AA\ implies a mean projected Doppler velocity of 10,000~$\pm$~500~km~s$^{-1}$. Looking red-ward to the feature centered about 6250~\AA, associating the minimum with \ion{Si}{2}~$\lambda$6355 implies a mean projected Doppler velocity of 5000~$\pm$~500~km~s$^{-1}$. This differs from that inferred from \ion{He}{1} by as much as 5000~km~s$^{-1}$. This suggests an association between \ion{Si}{2}~$\lambda$6355 and the minimum of the 6250~\AA\ feature is unlikely. 

Comparison between the day~$+$8 spectrum of SN~1984L and the day~$-$36 spectrum of the SN Ib 6p5Ax1 model from \citet{Dessart15} reveals moderate agreement. However, this particular benchmark model is not well-matched enough to invoke an unproven detection of \ion{Si}{2}~$\lambda$6355. Specifically, the projected Doppler velocities of the model and observations are not in sync. A visual comparison to the subsequent 6p5Ax1 model spectra (top-right panel of Figure~\ref{Fig:dessart3}) does not also reveal a more accurate match; i.e. the relative strength of the \ion{He}{1} signatures improve, but the projected Doppler velocities remain too low. We emphasize that the 6p5Ax1 model has not been built for SN~1984L, yet this model still does not provide accurate predictions for any SN~Ib shown in Figure~\ref{Fig:dessart3}. This may be due, in part, to signatures of \ion{Si}{2}~$\lambda$6355 that are too strong for the model.

There does exist fair visual agreement between the 6170~\AA\ feature in the SN~Ic~1987M and the 9000~km~s$^{-1}$~\ion{Si}{2} profile produced by \citet{Dessart12} for SN~Ib~2008D; i.e., in our Figure~\ref{Fig:dessart3}, overlap with the computed profile of \ion{Si}{2} is arguably convincing. However, a composition that is contaminated by trace hydrogen is difficult to rule out since the epoch of the only available spectrum ($\sim$~day~$+$9) is not early enough to disaffirm tentative detections of weak signatures from hydrogen and helium atoms \citep{Jeffery87M91}.

\begin{figure}
\centerline{\includegraphics*[scale=0.3, trim = 0mm 0mm 50mm 40mm]{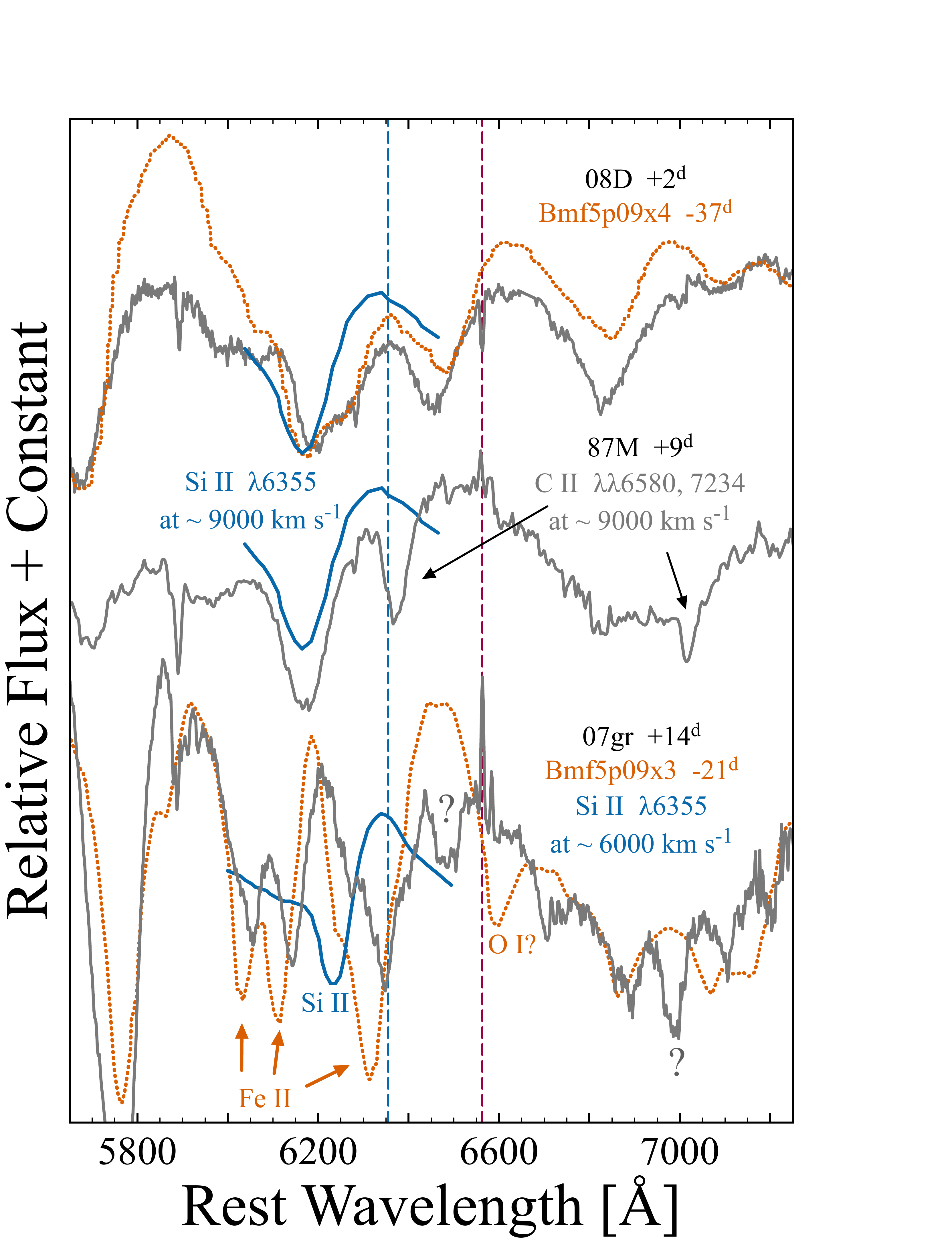}}
\caption{Model spectra from \citet{Dessart12} (in orange) compared to the SN~Ib~2008D and the SN~Ic~2007gr. The blue \ion{Si}{2} profiles computed for SN~2007gr and SN~2008D correspond to mean projected Doppler velocities of $\sim$~6000~km~s$^{-1}$ and 9000~km~s$^{-1}$, respectively.}
\label{Fig:08D07gr}
\end{figure}

For the SN~Ic~2007gr, which is plotted in Figures~\ref{Fig:dessart3}~and~\ref{Fig:08D07gr}, \citet{Dessart12} draw association between the feature centered about 6350~\AA\ as a composite of \ion{Si}{2} on the blue wing forming with a line velocity of $\sim$~6000~km~s$^{-1}$ and \ion{Fe}{2} near the minimum and red-ward. For the full fit, the day~$-$32 Bmf5p09x3 model spectrum at these wavelengths is promising when compared to the day~$+$13 spectrum of SN~2007gr. However, given the overall blue-ward displacement of the model compared to observations, we interpret the corresponding line velocities as being too high. 

Validation of this model for SN~2007gr could benefit, e.g., by a modest reduction of the kinetic energy released. (This was done by \citealt{Branch02} who used \texttt{SYNOW} to interpret and measure \ion{Fe}{2} lines in the SN~Ic~1990B.) However, the \ion{Si}{2} would still remain too blue to confidently associate \ion{Si}{2} with the absorption minimum of the 6350~\AA\ feature. If the computed spectrum were well-centered about this 6350~\AA\ feature, we suspect the absorption minima between the observed and synthesized spectrum would be significantly offset at other wavelengths. 

In the bottom of Figure~\ref{Fig:08D07gr}, one can also see that the computed spectrum produces a \ion{Si}{2}~$\lambda$5972 feature that is not observed for SN~2007gr in this region, which is possibly contaminated by narrow \ion{Na}{1}. In addition, there is less of a ``\ion{Si}{2}'' notch on the blue wing of the larger 6350~\AA\ feature to constrain the model when the observed spectrum is corrected for telluric features (cf. \citealt{Modjaz14}). Subsequently, we find the detection of \ion{Si}{2}~$\lambda$6355 in SN~2007gr to be either unlikely and therefore not representative of ``\ion{Si}{2} features'' of SN~Ib/Ic in general, or contribution from \ion{Si}{2}~$\lambda$6355 is minimal for these kinds of core-collapse events.

\subsection{Super-luminous Supernovae}

The same distinctions used for classical types I and II are also used for super-luminous supernovae. SLSN~II show conspicuous signatures of hydrogen, however the hydrogen feature is so far like that of SN~IIn \citep{Dessart15SLSN}, i.e. narrow emission signatures of hydrogen, e.g., SN~2006gy \citep{Smith07} and 2008am \citep{Chatzopoulos11}.

For SLSN~I, a spectral feature in the vicinity of 6000~$-$~6400~\AA\  is either truly absent (or faint under noise), or a large and likely blended P~Cygni profile is present. SLSN~I spectra with a conspicuous 6250~\AA\ feature, i.e. similar to those observed for SN~2007bi, LSQ12dlf, and iPTF13ehe, are frequently declared to be without a hydrogen feature, or even so far as ``hydrogen-free'' \citep{Galyam09,Nicholl15,Yan15}. 

However, recall from Fig.~1 of \citet{Filippenko97} where the 6250~\AA\ feature in SN~Ib/Ic spectra has been imprecisely labeled to signify an identification of ``weak \ion{Si}{2}'' despite a conflicting definition of no silicon detected within \S2.1 of \citet{Filippenko97}. Under the assumption that ``type~I means no hydrogen,'' an alternative interpretation of ``weak \ion{Si}{2}~$\lambda$6355'' is often applied to the relatively large 6250~\AA\ P~Cygni profile in the spectra of some SLSN~I and all BL-Ic.

Hence, if SN~Ib/Ic and SLSN~I shared the same ``weak silicon'' feature, then perhaps one would expect to see other lines of \ion{Si}{2} in the spectra of BL-Ic and SLSN~I where the feature is much stronger. Thus far, however, no strong evidence favoring a detection of other lines of silicon, e.g. \ion{Si}{2}~$\lambda\lambda$3858, 5972, has surfaced, while the low signal-to-noise ratios and low follow-up frequencies for these events prohibit more detailed investigations.

\begin{figure*}
\centerline{\includegraphics*[scale=0.45, trim = 0mm 0mm 30mm 0mm]{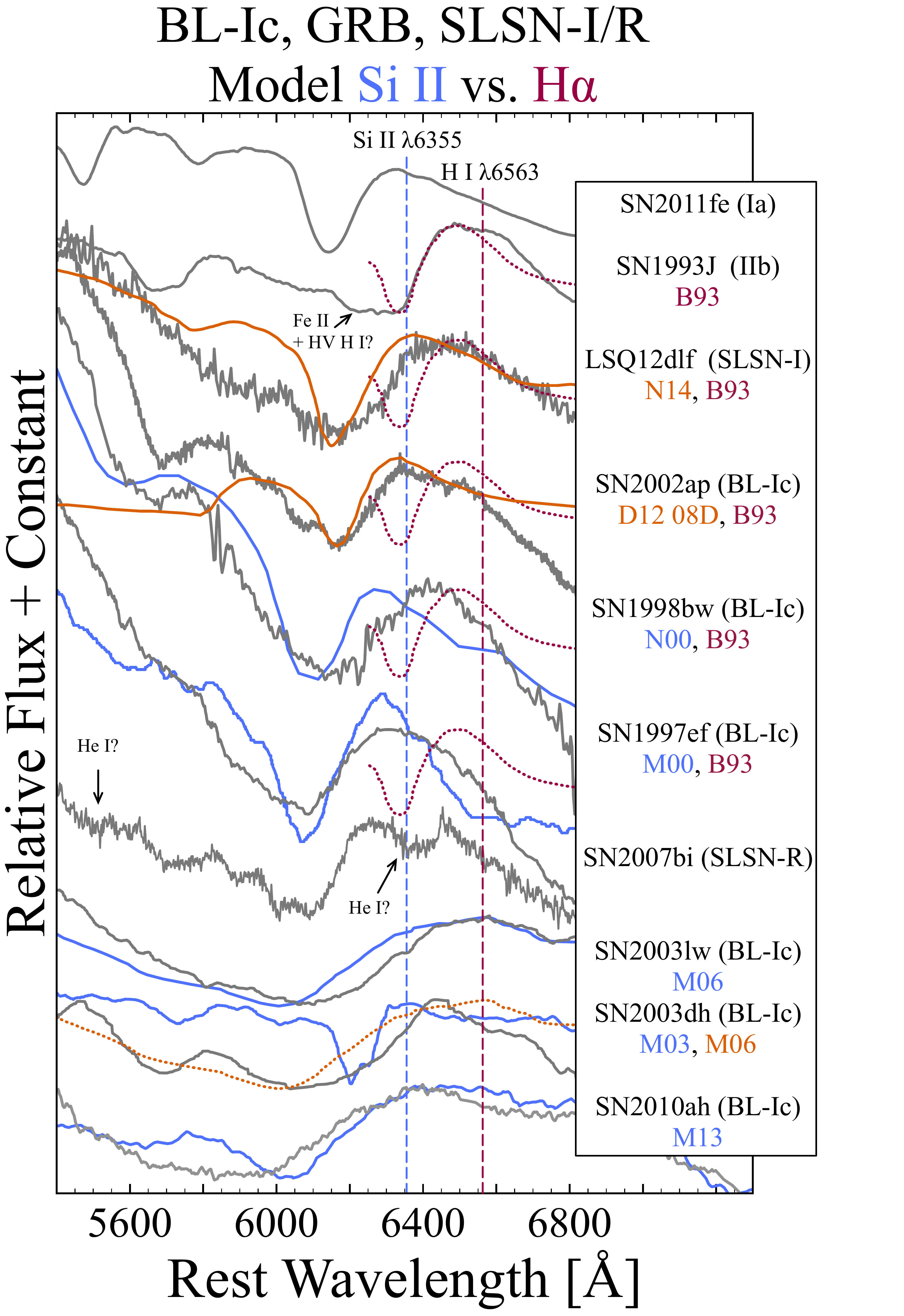}}
\caption{Comparisons between computed spectra and a selection of BL-Ic, SLSN, SN~1993J, and SN~2011fe. Spectrum references: SN~1993J \citep{Barbon95}; SN~1997ef \citep{Modjaz14}; SN~1998bw \citep{Patat01}; SN~2002ap \citep{Modjaz14}; SN~2003dh \citep{Deng05}; SN~2003lw \citep{Mazzali06}; SN~2007bi \citep{Galyam09}; SN~2010ah \citep{Corsi11}; SN~2011fe \citep{Pereira13}; LSQ12dlf \citep{Nicholl15}. Model references: (B93) \citet{Baron93}; (N00) \citet{Iwamoto98,Nomoto00}; (M00) \citet{Mazzali97ef00}; (M03) \citet{Mazzali03}; (M06) \citet{Mazzali06}; (D12~08D) \ion{Si}{2} profile from a computed spectrum by \citet{Dessart12}, initially compared to the SN~Ib~2008D; (M13) \citet{Mazzali13}; (N14) \citet{Nicholl14}.}
\label{Fig:SLSNsilicon}
\end{figure*}

In the following subsections we focus our attention on Figure~\ref{Fig:SLSNsilicon}, where we have plotted comparisons between computed and observed spectra of BL-Ic and SLSN~I found throughout the literature. Inconsistencies of models that produce signatures of \ion{Si}{2}~$\lambda$6355 are discussed in \S3.3.1, and we again highlight spectral influence from H$\alpha$, or ejecta contaminated by unburned hydrogen, as a promising alternative in \S3.3.2. 

\subsubsection{\ion{Si}{2}~$\lambda$6355 as a candidate identification}

Close inspection of Figure~\ref{Fig:SLSNsilicon} reveals an insufficient match between models that produce strong signatures of \ion{Si}{2}~$\lambda$6355 and observations of BL-Ic and SLSN. A mismatch of ``too blue'' in wavelength is observed for a majority of these fully-stripped C$+$O models from \citet{Iwamoto94}, and either without any improvement in subsequent adaptations to BL-Ic and SLSN, or with the same conclusion of a model with a relatively shallower density profile compared to the observations.

For the BL-Ic~1998bw the emission component produced by the model near 6500~\AA\ is nearly 100~\AA\ too blue to match observations without considerable absorption via {\it ad hoc} prescriptions of \ion{C}{2}, \ion{Ne}{1}, and \ion{Fe}{2}. However, if the model were improved near 6200~\AA\ with \ion{C}{2} and \ion{Ne}{1}, these ions would be at odds with the data at other wavelengths (cf. \citealt{Elmhamdi06,Elmhamdi07}). Alternatively, and using \texttt{SYNAPPS}, one could over-fit the region near 6300~\AA\ with \ion{C}{2} and \ion{Fe}{2} to secure a plausible match. Ultimately, however, the fit would suffer at bluer wavelengths, as shown recently by \citet{Toy15} for SN~2013dx, which is similarly associated with a gamma-ray burst event.

As for BL-Ic 2002ap \citep{Foley03}, the 9000~km~s$^{-1}$ \ion{Si}{2} profile from \citet{Dessart12} reveals similar problems of ``too blue'' without an {\it ad hoc} prescription of adjacent species that are not immediately detected elsewhere. In particular, the discrepancy between the curvature of the computed profile dominated by \ion{Si}{2}~$\lambda$6355 and that from observations indicates that \ion{Si}{2}~$\lambda$6355 cannot be directly associated with the minimum of the 6250~\AA\ feature in BL-Ic 1998bw and 2002ap. Similar conclusions apply to so-called ``Hypernovae'', such as SN~2003hd \citep{Deng05} and 2003lw \citep{Mazzali06}. In addition, the spectrum computed for SN~2003dh conflicts with observations throughout, including wavelength regions not shown in our Figure~\ref{Fig:SLSNsilicon}. 

Comparison between the the March 7 spectrum of SN~2010ah (PTF10bzf) and the computed spectrum of \citet{Mazzali13} is shown at the bottom of Figure~\ref{Fig:SLSNsilicon}. For this BL-Ic, \citet{Mazzali13} utilize rescaled CO138 models of \citet{Iwamoto98} to interpret spectral signatures. Here again, we see the model does not provide adequate predictions. However, unlike previous comparisons to BL-Ic and SLSN, this is the first time in Figure~\ref{Fig:SLSNsilicon} that we see a computed spectrum under-shoot the observed 6050~\AA\ feature. There are at least two avenues for addressing the discrepancy between a profile dominated by \ion{Si}{2}~$\lambda$6355 and observations of SN~2010ah. 

\begin{itemize}

\item [-] The blue wing of the 6150~\AA\ feature in the model conflicts with the observations, and similar to what is seen for SN~1997ef, 1998bw, 2002ap, 2003dh, and LSQ12dlf. One could posit that the model predicts velocities that are not as high as the observations, and that a simple scaling of kinetic energies, or a shallower ejecta density profile, would address the mismatch, thereby shifting the computed spectrum blue-ward. This is marginally true within the span of wavelengths plotted in Figure~\ref{Fig:SLSNsilicon}. However, such a correction for features near 6000~$-$~6400~\AA\ would create additional mismatch already seen for \ion{Ca}{2} and \ion{Fe}{2} features at other wavelengths (shown in Fig~6. of \citealt{Mazzali13}).  

\item[-] Alternatively, one could argue the mismatch between the model spectrum does not correspond to the epoch needed for an appropriate comparison to SN~2010ah. However, had follow-up observations been obtained, i.e. for more than the two spectra obtained for SN~2010ah, we suspect the CO138 model spectrum would appear ``too blue'' near 6250~\AA\ at an ostensibly appropriate epoch.

\end{itemize}

In Figure~\ref{Fig:SLSNsilicon}, the computed spectrum for the BL-Ic~1997ef \citep{Mazzali97ef00} looks promising near the absorption minimum of the 6200~\AA\ feature. However, viewed as a whole, it seems the parent models of CO21 \citep{Iwamoto94} and CO138H \citep{Iwamoto98} associated with the assortment of mismatched synthetic spectra are better suited for other supernovae yet to be observed. 

The ``type-R''~SN~2007bi has been suggested to be powered by radioactive decay through pair instability \citep{Galyam09}. This origin, however, has since been contested \citep{Moriya10,Yoshida11,Dessart13PISN,Chatzopoulos15}. Since a spectrum has not been computed for SN~2007bi, one can only guess at its composition by comparing SN~2007bi to other objects in Figure~~\ref{Fig:SLSNsilicon}. Based on the similarity of 6000$-$6400~\AA\ features, the 6250~\AA\ feature for SN~2007bi is either not dominated by \ion{Si}{2}, or the evidence favoring a detection of \ion{Si}{2} is insufficient. A similarly fleeting ``identification'' of \ion{He}{1} (indicated by the arrows in Figure~\ref{Fig:SLSNsilicon}) might instead favor either contamination from H$\alpha$ near 6200~\AA, or a composition that is not necessarily hydrogen-free.

Without sufficient improvements from stripped-envelope models that produce signatures of \ion{Si}{2}~$\lambda$6355, \citet{Nicholl14,Nicholl15} recently utilized \texttt{SYNAPPS} to identify \ion{Si}{2} in the spectrum of another SLSN~I, LSQ12dlf (shown near the top of our Figure~\ref{Fig:SLSNsilicon}). The interpretation of \ion{Si}{2} as the 6250~\AA\ feature in LSQ12dlf is clearly inconsistent with the data from the red-most side of the emission component, through the absorption trough, and near the blue-most wing. Considering the observed feature slopes below the emission component of the proposed fit, which includes some \ion{Fe}{2}, the mismatch is unavoidable without significant contribution from \ion{Fe}{2} that, in the end, is difficult to constrain at other wavelengths. Moreover, and similarly for SN~2013dx \citep{Toy15}, the spectrum of LSQ12dlf is too noisy to assess the underlying structure of the composite feature near 6250~\AA.

\citet{Yan15} recently presented observations of the SLSN~I~iPTF13ehe that revealed strong emission in H$\alpha$ during late-nebular phases. \citet{Yan15} conclude this CSM material originates from the progenitor in the years prior to the explosion, and estimate a CSM mass of $\lesssim$~30~M$_{\odot}$.  Given the progenitor masses for some of these SLSN~I are thought to be 67$-$220~M$_{\odot}$, this would imply the progenitor's mass is relatively high, and possibly with trace amounts of hydrogen by the time of the explosion. (This of course depends on the episodes of mass loss throughout the progenitor's evolution.)

Without providing a computed spectrum that is sufficiently matched to iPTF13ehe, \citet{Yan15} also interpret the 6000$-$6400~\AA\ feature as predominately \ion{Si}{2}~$\lambda$6355. Given the evidence for interaction between the ejecta of this SLSN~I with material contaminated by hydrogen that was previously ejected (see also \citealt{Chugai06,danmil15}), should we then expect trace amounts of hydrogen ($\sim$~10$^{-3}$~M$_{\odot}$) to remain for massive stars, possibily as massive as $\sim$~125~M$_{\odot}$? 

More recently, \citet{Arcavi14,Arcavi15} presented observations of the rapidly-rising PTF10iam. \citet{Arcavi15} claim that the identification of its 6200~\AA\ feature can be either \ion{Si}{2}~$\lambda$6355 or H$\alpha$. However, an interpretation of ``either-or'' is not applicable to unproven identifications of composite features.

Curiously, \citet{Arcavi15} find a fair match between PTF10iam and the SN~II~1999em, apart from SN~II-like H-Balmer lines that are absent for PTF10iam. \citet{Arcavi15} also compare the day~$+$28 spectrum of PTF10iam to the day~$+$11 spectrum of the SN~Ia~1999ac, and claim that ``the SN Ia fit is able to match this feature as Si II, as well as other features in the spectrum, with the major difference being that PTF10iam has additional broad hydrogen emission lines.''

However, we find another major difference is that the 6200~\AA\ feature in PTF10iam is not well-matched to the 6135~\AA\ feature of SN~1999ac. In particular, the feature for SN~1999ac is noticeably too blue compared to that of PTF10iam. Furthermore, if the SN~Ia~1999ac can be used to identify the 6200~\AA\ feature of PTF10iam as \ion{Si}{2}~$\lambda$6355, while also ``matching other features'', why then is PTF10iam not classified as a SN~Ia interacting with a H-rich medium?\footnote{PTF10iam is not likely a SN~Ia because it does not look like a SN~Ia prior to day~$+$28. Yet an identification for its 6200~\AA\ feature is being interpreted as \ion{Si}{2}~$\lambda$6355 based on an unmatched comparison to the SN~Ia~1999ac, while an identification of \ion{Si}{2}~$\lambda$6355 is so far only definitely detected for SN~Ia \citep{Wheeler95,Filippenko97}.}

The spectra of PTF10iam are also noisy, and therefore do not have a well-defined minimum for the feature near 6200~\AA\ (see also \citealt{Greiner15,Leloudas15LSQ14mo}). However, for a 6200~\AA\ feature to be identified as \ion{Si}{2}~$\lambda$6355, this corresponds to a line velocity of $\sim$~8000~km~s$^{-1}$ ($\pm$~1000~km~s$^{-1}$). For the SN~Ia~1999ac spectrum that \citet{Arcavi15} claim to have matched to PTFiam's 6200~\AA\ feature, its \ion{Si}{2}~$\lambda$6355 line velocity is approximated to be $\sim$~11,000~km~s$^{-1}$. Consequently, it is unlikely that the 6200~\AA\ feature of PTF10iam can be solely attributed to \ion{Si}{2}~$\lambda$6355, and we suspect models that produce spectra dominated by \ion{Si}{2}~$\lambda$6355 will be unable to unequivocally match both the minimum and the blue and red wings of the 6200~\AA\ feature without producing inconsistencies elsewhere. 

This reliance on \ion{Si}{2}~$\lambda$6355 as a candidate identification reveals an ongoing problem. So far, there has been no observational precedent for lone signatures of \ion{Si}{2}~$\lambda$6355 outside of SN~Ia \citep{Wheeler95,Filippenko97}, nor has an accurate search of additional signatures of \ion{Si}{2} been carried out. Yet the common wisdom for SN~Ib, Ic, BL-Ic, and SLSN continues to default to a subjective and inconsistent identification of \ion{Si}{2}~$\lambda$6355. It is therefore a wonder why \ion{Si}{2}~$\lambda$6355 has been so strongly favored as an interpretation (cf. Fig.~1 of \citealt{Filippenko97} and Fig.~2 of \citealt{Modjaz14}). A more likely alternative identification is not necessarily H$\alpha$, much less H$\alpha$ alone. Rather, a 6200~\AA\ feature that is influenced by trace amounts of ejected hydrogen would be supported by the same observations of PTF10iam purportedly interacting with a H-rich medium near maximum light \citep{Arcavi15}.

\subsubsection{\ion{H}{1}~$\lambda$6563 as a candidate identification}

\begin{figure}
\centerline{\includegraphics*[scale=0.42, trim = 100mm 0mm 120mm 10mm]{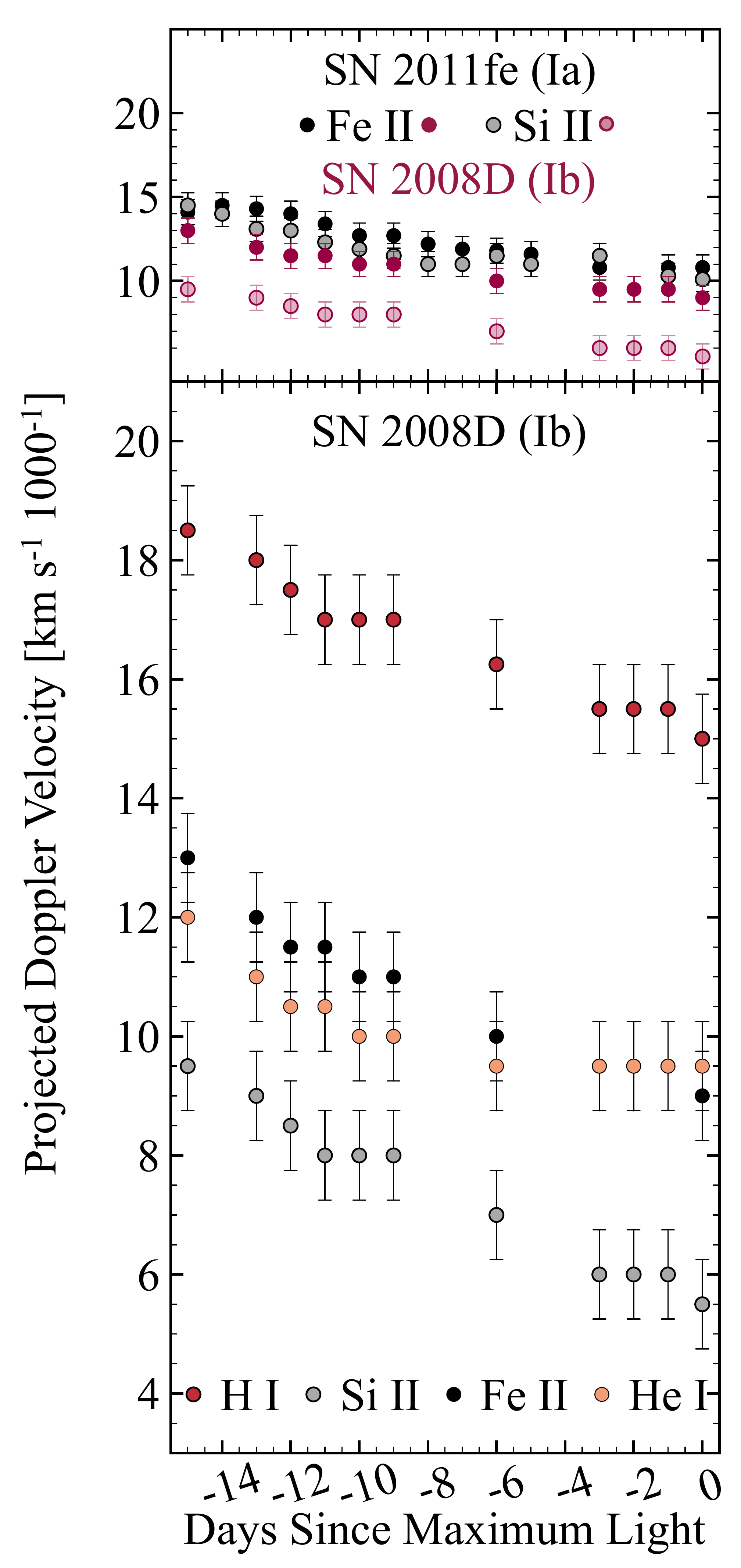}}
\caption{Estimated line velocities of SN~Ib~2008D from its absorption minima. Vertical error bars shown for \ion{Si}{2} and \ion{Fe}{2} represent a minimum error of $\pm$~750 km~s$^{-1}$. The errors could be as high as 2000~km~s$^{-1}$. Shown above are values obtained for the SN~Ia~2011fe using \texttt{SYNAPPS} \citep{Parrent12}.}
\label{Fig:08Ddopplers}
\end{figure}

Compared to H$\alpha$ in SN~II spectra, SN~IIb show weaker, albeit typical, signatures of hydrogen alongside conspicuous helium lines \citep{Filippenko88}. The H$\alpha$ profile for SN~IIb is also generally broader than that for SN~II because the mass of the remaining hydrogen is greater for SN~II than SN~IIb. Consequently, the relative weakness of H$\alpha$ in SN~IIb spectra enables adjacent blending from species such as \ion{Si}{2}, \ion{Fe}{2}, and possibly higher velocity \ion{H}{1} to broaden a spectral feature that would otherwise be solely attributed to H$\alpha$ (\citealt{Danmil1311ei} and references therein). 

Considering the similarity of all BL-Ic and select SLSN~I/R plotted in Figure~\ref{Fig:SLSNsilicon}, we find contamination by H$\alpha$ to be a plausible solution. One caveat is that the classically defined rest-wavelength P~Cygni emission component of the observed feature is not centered about a rest-wavelength of 6563~\AA. For SN~II, \citet{Anderson14} have shown the blue-shifted emission of H$\alpha$ is sensitive to the ejecta density profile, particularly when the radial falloff is steep. Blue-shifted emission components in H$\alpha$ have also been observed for SN~IIb, e.g., the well-observed SN 1993J \citep{Baron93}. Therefore, another solution that would support an interpretation of hydrogen for BL-Ic and select SLSN~I/R is if the underlying emission component of H$\alpha$ is blue-shifted along with the corresponding absorption minimum.  


\subsection{Measures of projected Doppler velocity}


Estimates of so-called detachment velocities for hydrogen are either accurate from first principles, or direct measurements are again spoiled by effects of radiation transport in that the absorption minimum does not accurately estimate mean projected Doppler velocities for H$\alpha$ (cf. \citealt{Branch77}). Under the assumption of either localized line formation, or multiple resonance line scattering (effectively \texttt{SYNAPPS}), line velocities inferred from absorption minima may over-shoot the optimum position of these ostensibly found hydrogen-deficient regions of ejecta. If faint H$\alpha$ in SN~Ib, Ic, BL-Ic, and SLSN~I/R behaves similarly compared to SN~II and IIb, then trace amounts of detectable hydrogen would indicate high projected Doppler velocities in spite of being relatively closer to the photospheric line forming region (cf. \citealt{Anderson14}). 

By contrast, the red-shift needed to rectify ill-matched signatures of \ion{Si}{2}~$\lambda$6355 is on the order of 200~\AA. The effect that opposing interpretations of \ion{Si}{2}~$\lambda$6355 versus H$\alpha$ has on estimates of line velocities is shown in Figure~\ref{Fig:08Ddopplers} for the SN~Ib~2008D. Assuming one line dominates the 6250~\AA\ feature, the error would be on the order of $\delta$$v$~$\sim$~9000~km~s$^{-1}$ between \ion{Si}{2}~$\lambda$6355 and H$\alpha$; i.e., noticeably large when spectroscopic resolutions of blended features are on the order of~$\gtrsim$~750~km~s$^{-1}$ for these classes of SN~I. Compared to the SN~Ia~2011fe, where signatures of both \ion{Si}{2} and \ion{Fe}{2} can be used to trace respective line velocities, the same conclusion is not reached when considering the SN~Ib~2008D.

\begin{figure}
\centerline{\includegraphics*[scale=0.4, trim = 85mm 45mm 130mm 30mm]{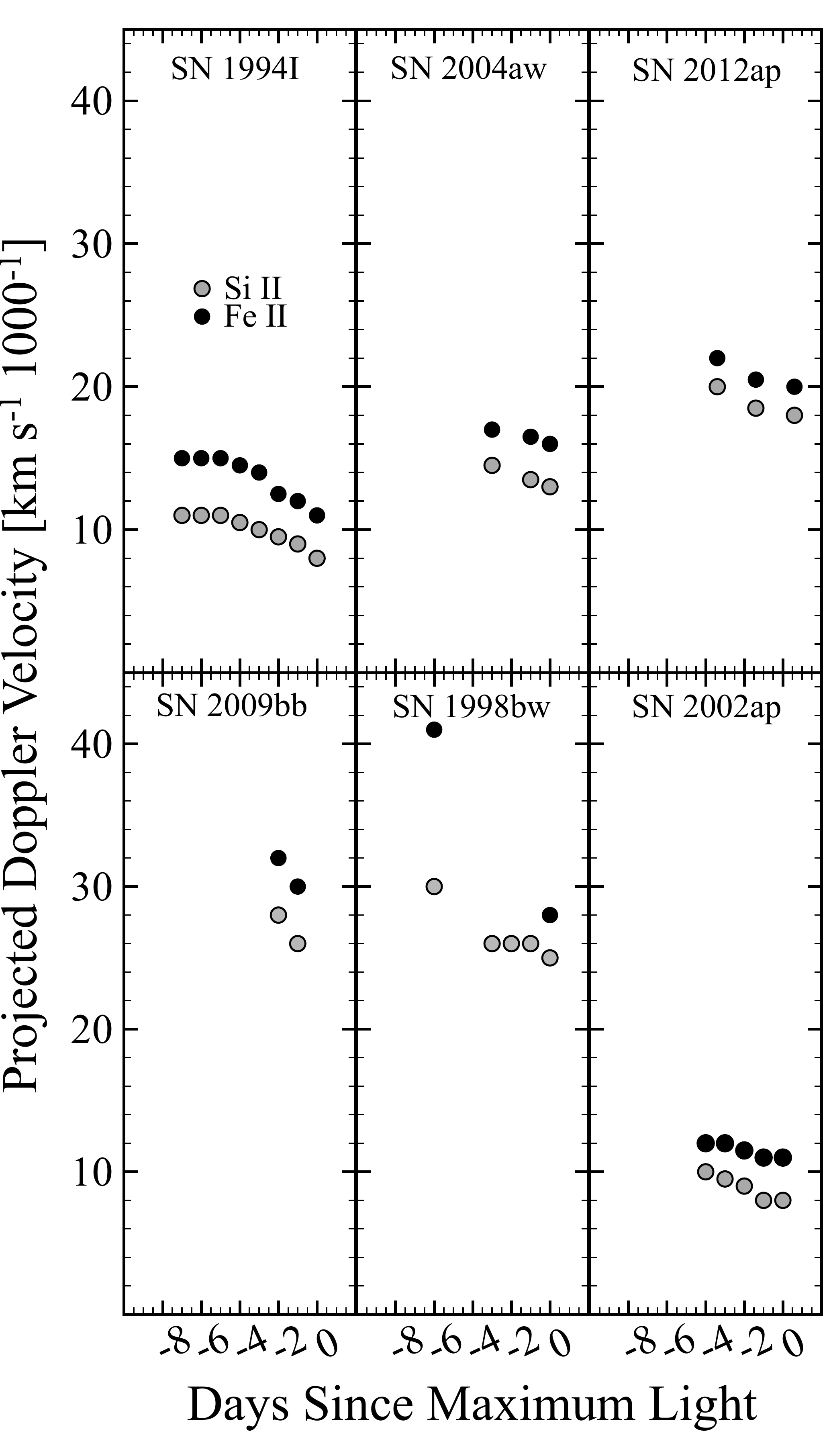}}
\caption{Estimated line velocities of 6000~\AA\ features in SLSN and BL-Ic spectra. The grey points denote values obtained under an assumption that the absorption minimum is identified with \ion{Si}{2}~$\lambda$6355. Estimates of line velocities for \ion{Fe}{2} are the points shown in black. For some of the available observations, we excluded points that did not have either an obvious minimum or underlying structure.}
\label{Fig:helium}
\end{figure}

Similarly, line velocities of \ion{Fe}{2} are also noticeably higher than that of \ion{Si}{2}~$\lambda$6355 for SN~1994I, 1998bw, 2002ap, 2004aw, 2009bb, and 2012ap, shown in Figure~\ref{Fig:helium} (see also Table~2 of \citealt{Stathakis00}). This trend continues into post-maximum epochs that are not shown, and indicates that if a signature of \ion{Si}{2}~$\lambda$6355 is shaping the spectrum, then it is weak and contributing to the blue-most wing of 6000~\AA\ absorption features (instead of dominating an entire profile, much less the minimum).

\section{Summary and Conclusions}

In this work we examined publicly unavailable model spectra from both ``hydrogen-poor'' (M$_{H}$~$\sim$~10$^{-3}$~$-$~10$^{-2}$~M$_{\odot}$) and ``hydrogen-depleted'' (M$_{H}$~$\sim$~10$^{-6}$~M$_{\odot}$) compositions and compared them to observations of multiple type~I supernova subclasses. Apart from natural uncertainties from line blending, and regardless of the spectrum synthesizer used, all 1-dimensional model spectra intended for thermonuclear SN~Ia are well-matched (Figure~\ref{Fig:SNIasilicon}). Since the evidence favoring a detection of \ion{Si}{2} $\lambda$6355 is overwhelming, direct association between \ion{Si}{2}~$\lambda$6355 and the minimum of 6150~\AA\ features can be made in practice. 

However for SN~Ib, Ic, BL-Ic, and SLSN~I/R, the evidence favoring detections of \ion{Si}{2}~$\lambda$6355 spans few events. This implies the direct association between \ion{Si}{2}~$\lambda$6355 and the minimum of 6000$-$6400~\AA\ feature is not accurate. In addition, the observational evidence that would otherwise support an interpretation of \ion{Si}{2}~$\lambda$6355, e.g., signatures from lines other than $\lambda$6355, is as fleeting as any weak evidence that would similarly confirm the presence of trace hydrogen through H$\beta$ (Figure~\ref{Fig:Hbeta}).\footnote{Although section 2.1 of \citet{Filippenko97} claims \ion{Si}{2}~$\lambda$6355 is detected for SN Ia, but not for SN~Ib/Ic, the labels of \ion{Si}{2}~$\lambda$6355 against SN~Ia, Ib, and Ic spectra in Fig.~1 of \citet{Filippenko97} are at odds with this statement. As a result, references made to this review have sided with either the labels of Fig.~1 or the notions expressed within the body of the text.}

Among the available spectra stemming from models of stripped-envelope supernovae, and compared to observed 6000$-$6400 spectral features of SN~Ib, Ic, BL-Ic, and SLSN, we find a consistent blue-ward mismatch in wavelength, in part, due to signatures of \ion{Si}{2}~$\lambda$6355 that are too strong. Subsequent explanations for this phenomenon have included a shallow density profile for the corresponding models (\citealt{Hachinger12} and references therein), or a prescription that is out of sync with the time-series observations (see our \S3.3.1). However, these notions are counter to both the successful interpretation of \ion{Si}{2} for SN~Ia (Figure~\ref{Fig:SNIasilicon}), and the categorical shortcomings of stripped-envelope models that produce signatures of \ion{Si}{2} that are too strong for pre-maximum observations of SN~Ib, Ic, BL-Ic, and SLSN.

The inability of stripped-envelope models to accurately reproduce the shape of a given spectral feature could be attributed to many potential issues with velocities, composition, excitation/ionization, and assumptions of spherical symmetry in the model, and would not necessarily imply that a given model is fundamentally wrong (or right). However, if both photometry and time-series spectra of a given event are to be thought of as sensitive to the radiation field and properties of the ejecta, then a model that is only approximately consistent with one of these pieces of information precludes firm conclusions about the nature of the spectral sequence being studied. 

Contribution from species with line signatures near 6250~\AA, such as \ion{C}{2} and \ion{Fe}{2}, may help to both prevent a poor match for a model and refine prescriptions for empirical measurements (cf. \citealt{Dessart15}). This solution, however, will require simultaneous agreement between the observed and computed photometry since this is so far not the case when an interpretation of \ion{Si}{2} with \ion{Fe}{2} has been put forward  (see \S3.2). 

Furthermore, today's spectrum-limited observations do not provide the necessary constraints for invoking detections of \ion{C}{2} through both $\lambda$6580 and $\lambda$7234 counterparts. This underscores the need for additional surveys that embark on maximizing follow-up frequency and signal-to-noise ratios throughout the evolution of the spectrum.

Among the most promising matches with stripped-envelope core-collapse models, two ``realistic'' model spectra can be described as being in fair agreement with two observations from two different objects; namely the hydrogen-poor (M$_{H}$~$\sim$~10$^{-3}$~M$_{\odot}$) and compositionally mixed SN Ib model Bfm4p41x4 \citep{Dessart12} compared to the SN~Ib~1999dn, and the hydrogen-depleted (M$_{H}$~$\sim$~10$^{-6}$~M$_{\odot}$) SN Ib model Bmf5p09x4 \citep{Dessart12} compared to the SN~Ib~2004gq. Even so, the rise-times for these models are not in good agreement with observations for these and most other SN Ib/Ic (cf. \citealt{Wheeler15}), which brings into question the validity of the models altogether.


One might expect detailed modeling through \texttt{PHOENIX} and \texttt{CMFGEN}, both of which assume spherical symmetry for post-processing, to show some improvement at wavelengths red-ward of 6250~\AA\ for hydrogen-depleted compositions since these synthesizers at least consider a larger volume of line formation than those methods that assume a sharp photosphere. However, such a red-ward fix does not appear to reveal itself in either the \texttt{PHOENIX} or \texttt{CMFGEN} synthetic spectra displayed in our Figures~\ref{Fig:dessart1},~\ref{Fig:dessart2},~\ref{Fig:dessart3},~and~\ref{Fig:08D07gr}. In addition, an apparent absence of 200~\AA\ line-shifts toward redder wavelengths for other spectral features appears to suggest that considerations of 3-dimensional radiation transport would not conspire to enable the minimum of a \ion{Si}{2} $\lambda$6355 profile to coincide with the observed minimum near 6250~\AA. 


Thus, an absence of faint H$\alpha$ originating from trace hydrogen ($\sim$~10$^{-3}$~M$_{\odot}$) is not simply a given for these ``type~I'' supernovae, while the common wisdom has been to judge the quality of a computed spectrum from a fully-stripped model by its [in]ability to match an unproven signature of \ion{Si}{2}~$\lambda$6355 within the feature centered about 6250~\AA. Moreover, observations of would-be conspicuous signatures of hydrogen in the spectra of BL-Ic and SLSN~I suggest that the broad emission component of H$\alpha$ undergoes a significant blue-ward shift relative to the rest wavelength of 6563~\AA, which is an effect known to arise from steep density-gradients in the outermost ejecta of SN~II in general \citep{Baron95,Anderson14}.  

In the literature there are at least three promising models with computed spectra where $\sim$~10$^{-3}$~M$_{\odot}$ of hydrogen is present within the outermost layers of ejecta, and the times-series spectra for only one of these models is ``H$\alpha$-negative''. These include the SN~IIb 3p65Ax1 \citep{Dessart15}, the aforementioned H$\alpha$-negative SN~Ib Bmf4p41x4 \citep{Dessart12}, and the calculations done by \citet{James10} who previously showed that H$\alpha$ cannot be ruled out assuming extended regions of trace amounts of hydrogen for the SN~Ib~1999dn.  Hence, whether or not the computed spectrum for a hydrogen-poor event is H$\alpha$-positive, the mass of hydrogen in the outermost layers of some caught-early SN~Ib, and therefore the progenitor, need not fall below $\sim$~10$^{-3}$~M$_{\odot}$ for the model to remain consistent with 6250~\AA\ spectral features (\S3.2.1).



Depending on the sample of objects utilized to infer the progenitor and mass-loss history of ``hydrogen-poor'' supernovae (e.g., IIb/Ib), relationships may exist between properties of the light curves and the extent of trace hydrogen throughout the ejecta. Given a suite of improved model spectra, faint detections of H$\alpha$, as well as non-detections stemming from trace hydrogen in the ejecta, might also be used to gain insights into the distribution of hydrogen near the surface of the progenitor star, or binary product, at the time of explosion. 

\acknowledgements

This work was made possible by contributions to the Supernova Spectrum Archive \citep{Richardson01} and the Weizmann Interactive Supernova data REPository (WISeREP \citealt{WISEREP}), as well as David Bishop's Latest Supernovae page \citep{Galyam13}. This work has also made use of the Sternberg Astronomical Institute Supernova Light Curve Catalogue, which is supported by grants of ``Scientific Schools of Russia (3458.2010.2)'', RFBR (10-02-00249).

All model spectra, as well as most of the data that were not available on WISeREP at the time of our initial study were obtained with the help of the graph digitizer software, GraphClick.\footnote{The full software is available at http://www.arizona-software.ch/graphclick/.} (Digitizing one published spectrum of a supernova by hand takes about as long as it does to obtain the original spectrum with a CCD chip; $\sim$~20$-$50 minutes depending on the signal-to-noise ratio.)

JTP is indebted to David Branch and J. Craig Wheeler for extensive discussions and valuable advice during the preparation of this manuscript. This work also benefitted from discussions with Rafaella Margutti, Maria Drout, Brian Friesen, Federica Bianco, and Isaac Shivvers, and additional comments from Gast\'{o}n Folatelli. 

Finally, we wish to thank our anonymous referee for taking the time to give constructive feedback. 

\bibliographystyle{apj}
\bibliography{jparrent_bib}{}

\begin{thebibliography}{125}
\expandafter\ifx\csname natexlab\endcsname\relax\def\natexlab#1{#1}\fi

\bibitem[{{Anderson} {et~al.}(2014){Anderson}, {Dessart}, {Gutierrez}, {Hamuy},
  {Morrell}, {Phillips}, {Folatelli}, {Stritzinger}, {Freedman},
  {Gonz{\'a}lez-Gait{\'a}n}, {McCarthy}, {Suntzeff}, \&
  {Thomas-Osip}}]{Anderson14}
{Anderson}, J.~P., {et~al.} 2014, \mnras, 441, 671

\bibitem[{{Arcavi} {et~al.}(2014){Arcavi}, {Gal-Yam}, {Sullivan}, {Pan},
  {Cenko}, {Horesh}, {Ofek}, {De Cia}, {Yan}, {Yang}, {Howell}, {Tal},
  {Kulkarni}, {Tendulkar}, {Tang}, {Xu}, {Sternberg}, {Cohen}, {Bloom},
  {Nugent}, {Kasliwal}, {Perley}, {Quimby}, {Miller}, {Theissen}, \&
  {Laher}}]{Arcavi14}
{Arcavi}, I., {et~al.} 2014, ApJ, 793, 38

\bibitem[{{Arcavi} {et~al.}(2015){Arcavi}, {Wolf}, {Howell}, {Bildsten},
  {Leloudas}, {Hardin}, {Prajs}, {Perley}, {Svirski}, {Gal-Yam}, {Katz},
  {McCully}, {Cenko}, {Lidman}, {Sullivan}, {Valenti}, {Astier}, {Balland},
  {Carlberg}, {Conley}, {Fouchez}, {Guy}, {Pain}, {Palanque-Delabrouille},
  {Perrett}, {Pritchet}, {Regnault}, {Rich}, \& {Ruhlmann-Kleider}}]{Arcavi15}
---. 2015, arXiv:1511.00704

\bibitem[{{Barbon} {et~al.}(1995){Barbon}, {Benetti}, {Cappellaro}, {Patat},
  {Turatto}, \& {Iijima}}]{Barbon95}
{Barbon}, R., {Benetti}, S., {Cappellaro}, E., {Patat}, F., {Turatto}, M., \&
  {Iijima}, T. 1995, \aaps, 110, 513

\bibitem[{{Baron} {et~al.}(1999){Baron}, {Branch}, {Hauschildt}, {Filippenko},
  \& {Kirshner}}]{Baron99}
{Baron}, E., {Branch}, D., {Hauschildt}, P.~H., {Filippenko}, A.~V., \&
  {Kirshner}, R.~P. 1999, ApJ, 527, 739

\bibitem[{{Baron} {et~al.}(1993){Baron}, {Hauschildt}, {Branch}, {Wagner},
  {Austin}, {Filippenko}, \& {Matheson}}]{Baron93}
{Baron}, E., {Hauschildt}, P.~H., {Branch}, D., {Wagner}, R.~M., {Austin},
  S.~J., {Filippenko}, A.~V., \& {Matheson}, T. 1993, ApJL, 416, L21

\bibitem[{{Baron} {et~al.}(1995){Baron}, {Hauschildt}, \&
  {Mezzacappa}}]{Baron95}
{Baron}, E., {Hauschildt}, P.~H., \& {Mezzacappa}, A. 1995,
  arXiv:astro-ph/9511081

\bibitem[{{Baron} {et~al.}(1996){Baron}, {Hauschildt}, {Nugent}, \&
  {Branch}}]{Baron96}
{Baron}, E., {Hauschildt}, P.~H., {Nugent}, P., \& {Branch}, D. 1996, MNRAS,
  283, 297

\bibitem[{{Benetti} {et~al.}(2004){Benetti}, {Meikle}, {Stehle}, {Altavilla},
  {Desidera}, {Folatelli}, {Goobar}, {Mattila}, {Mendez}, {Navasardyan},
  {Pastorello}, {Patat}, {Riello}, {Ruiz-Lapuente}, {Tsvetkov}, {Turatto},
  {Mazzali}, \& {Hillebrandt}}]{Benetti04}
{Benetti}, S., {et~al.} 2004, MNRAS, 348, 261

\bibitem[{{Blondin} {et~al.}(2015){Blondin}, {Dessart}, \&
  {Hillier}}]{Blondin15}
{Blondin}, S., {Dessart}, L., \& {Hillier}, D.~J. 2015, \mnras, 448, 2766

\bibitem[{{Blondin} \& {Tonry}(2007)}]{Blondin07}
{Blondin}, S., \& {Tonry}, J.~L. 2007, \apj, 666, 1024

\bibitem[{{Branch}(1977)}]{Branch77}
{Branch}, D. 1977, MNRAS, 179, 401

\bibitem[{{Branch} {et~al.}(2005){Branch}, {Baron}, {Hall}, {Melakayil}, \&
  {Parrent}}]{Branch05}
{Branch}, D., {Baron}, E., {Hall}, N., {Melakayil}, M., \& {Parrent}, J. 2005,
  PASP, 117, 545

\bibitem[{{Branch} {et~al.}(2002){Branch}, {Benetti}, {Kasen}, {Baron},
  {Jeffery}, {Hatano}, {Stathakis}, {Filippenko}, {Matheson}, {Pastorello},
  {Altavilla}, {Cappellaro}, {Rizzi}, {Turatto}, {Li}, {Leonard}, \&
  {Shields}}]{Branch02}
{Branch}, D., {et~al.} 2002, ApJ, 566, 1005

\bibitem[{{Branch} {et~al.}(2006){Branch}, {Dang}, {Hall}, {Ketchum},
  {Melakayil}, {Parrent}, {Troxel}, {Casebeer}, {Jeffery}, \&
  {Baron}}]{Branch06}
---. 2006, PASP, 118, 560

\bibitem[{{Chakraborti} {et~al.}(2015){Chakraborti}, {Ray}, {Smith},
  {Margutti}, {Pooley}, {Bose}, {Sutaria}, {Chandra}, {Dwarkadas}, {Ryder}, \&
  {Maeda}}]{Chakraborti15}
{Chakraborti}, S., {et~al.} 2015, arXiv:1510.06025

\bibitem[{{Chatzopoulos} {et~al.}(2015){Chatzopoulos}, {van Rossum}, {Craig},
  {Whalen}, {Smidt}, \& {Wiggins}}]{Chatzopoulos15}
{Chatzopoulos}, E., {van Rossum}, D.~R., {Craig}, W.~J., {Whalen}, D.~J.,
  {Smidt}, J., \& {Wiggins}, B. 2015, \apj, 799, 18

\bibitem[{{Chatzopoulos} {et~al.}(2011){Chatzopoulos}, {Wheeler}, {Vinko},
  {Quimby}, {Robinson}, {Miller}, {Foley}, {Perley}, {Yuan}, {Akerlof}, \&
  {Bloom}}]{Chatzopoulos11}
{Chatzopoulos}, E., {et~al.} 2011, \apj, 729, 143

\bibitem[{{Childress} {et~al.}(2013){Childress}, {Scalzo}, {Sim}, {Tucker},
  {Yuan}, {Schmidt}, {Cenko}, {Silverman}, {Contreras}, {Hsiao}, {Phillips},
  {Morrell}, {Jha}, {McCully}, {Filippenko}, {Anderson}, {Benetti}, {Bufano},
  {de Jaeger}, {Forster}, {Gal-Yam}, {Le Guillou}, {Maguire}, {Maund},
  {Mazzali}, {Pignata}, {Smartt}, {Spyromilio}, {Sullivan}, {Taddia},
  {Valenti}, {Bayliss}, {Bessell}, {Blanc}, {Carson}, {Clubb}, {de Burgh-Day},
  {Desjardins}, {Fang}, {Fox}, {Gates}, {Ho}, {Keller}, {Kelly}, {Lidman},
  {Loaring}, {Mould}, {Owers}, {Ozbilgen}, {Pei}, {Pickering}, {Pracy}, {Rich},
  {Schaefer}, {Scott}, {Stritzinger}, {Vogt}, \& {Zhou}}]{Childress13}
{Childress}, M.~J., {et~al.} 2013, ApJ, 770, 29

\bibitem[{{Chugai} \& {Chevalier}(2006)}]{Chugai06}
{Chugai}, N.~N., \& {Chevalier}, R.~A. 2006, \apj, 641, 1051

\bibitem[{{Clocchiatti} {et~al.}(1996){Clocchiatti}, {Wheeler}, {Brotherton},
  {Cochran}, {Wills}, {Barker}, \& {Turatto}}]{Clocchiatti96}
{Clocchiatti}, A., {Wheeler}, J.~C., {Brotherton}, M.~S., {Cochran}, A.~L.,
  {Wills}, D., {Barker}, E.~S., \& {Turatto}, M. 1996, \apj, 462, 462

\bibitem[{{Corsi} {et~al.}(2011){Corsi}, {Ofek}, {Frail}, {Poznanski},
  {Arcavi}, {Gal-Yam}, {Kulkarni}, {Hurley}, {Mazzali}, {Howell}, {Kasliwal},
  {Green}, {Murray}, {Sullivan}, {Xu}, {Ben-ami}, {Bloom}, {Cenko}, {Law},
  {Nugent}, {Quimby}, {Pal'shin}, {Cummings}, {Connaughton}, {Yamaoka}, {Rau},
  {Boynton}, {Mitrofanov}, \& {Goldsten}}]{Corsi11}
{Corsi}, A., {et~al.} 2011, \apj, 741, 76

\bibitem[{{Deng} {et~al.}(2005){Deng}, {Tominaga}, {Mazzali}, {Maeda}, \&
  {Nomoto}}]{Deng05}
{Deng}, J., {Tominaga}, N., {Mazzali}, P.~A., {Maeda}, K., \& {Nomoto}, K.
  2005, \apj, 624, 898

\bibitem[{{Deng} {et~al.}(2000){Deng}, {Qiu}, {Hu}, {Hatano}, \&
  {Branch}}]{Deng00}
{Deng}, J.~S., {Qiu}, Y.~L., {Hu}, J.~Y., {Hatano}, K., \& {Branch}, D. 2000,
  ApJ, 540, 452

\bibitem[{{Dessart} {et~al.}(2015{\natexlab{a}}){Dessart}, {Audit}, \&
  {Hillier}}]{Dessart15SLSN}
{Dessart}, L., {Audit}, E., \& {Hillier}, D.~J. 2015{\natexlab{a}}, MNRAS, 449,
  4304

\bibitem[{{Dessart} {et~al.}(2014{\natexlab{a}}){Dessart}, {Blondin},
  {Hillier}, \& {Khokhlov}}]{Dessart14models}
{Dessart}, L., {Blondin}, S., {Hillier}, D.~J., \& {Khokhlov}, A.
  2014{\natexlab{a}}, MNRAS, 441, 532

\bibitem[{{Dessart} {et~al.}(2014{\natexlab{b}}){Dessart}, {Hillier},
  {Blondin}, \& {Khokhlov}}]{Dessart14}
{Dessart}, L., {Hillier}, D.~J., {Blondin}, S., \& {Khokhlov}, A.
  2014{\natexlab{b}}, MNRAS, 439, 3114

\bibitem[{{Dessart} {et~al.}(2012){Dessart}, {Hillier}, {Li}, \&
  {Woosley}}]{Dessart12}
{Dessart}, L., {Hillier}, D.~J., {Li}, C., \& {Woosley}, S. 2012, MNRAS, 424,
  2139

\bibitem[{{Dessart} {et~al.}(2015{\natexlab{b}}){Dessart}, {Hillier},
  {Woosley}, {Livne}, {Waldman}, {Yoon}, \& {Langer}}]{Dessart15}
{Dessart}, L., {Hillier}, D.~J., {Woosley}, S., {Livne}, E., {Waldman}, R.,
  {Yoon}, S.-C., \& {Langer}, N. 2015{\natexlab{b}}, \mnras, 453, 2189

\bibitem[{{Dessart} {et~al.}(2013){Dessart}, {Waldman}, {Livne}, {Hillier}, \&
  {Blondin}}]{Dessart13PISN}
{Dessart}, L., {Waldman}, R., {Livne}, E., {Hillier}, D.~J., \& {Blondin}, S.
  2013, \mnras, 428, 3227

\bibitem[{{Doggett} \& {Branch}(1985)}]{Doggett85}
{Doggett}, J.~B., \& {Branch}, D. 1985, AJ, 90, 2303

\bibitem[{{Doull} \& {Baron}(2011)}]{Doull11}
{Doull}, B.~A., \& {Baron}, E. 2011, PASP, 123, 765

\bibitem[{{Drout} {et~al.}(2015){Drout}, {Milisavljevic}, {Parrent},
  {Margutti}, {Kamble}, {Soderberg}, {Challis}, {Chornock}, {Fong}, {Frank},
  {Gehrels}, {Graham}, {Hsiao}, {Itagaki}, {Kasliwal}, {Kirshner}, {Macomb},
  {Marion}, {Norris}, \& {Phillips}}]{Drout15}
{Drout}, M.~R., {et~al.} 2015, arXiv:1507.02694

\bibitem[{{Eldridge} {et~al.}(2013){Eldridge}, {Fraser}, {Smartt}, {Maund}, \&
  {Crockett}}]{Eldridge13}
{Eldridge}, J.~J., {Fraser}, M., {Smartt}, S.~J., {Maund}, J.~R., \&
  {Crockett}, R.~M. 2013, \mnras, 436, 774

\bibitem[{{Elmhamdi} {et~al.}(2007){Elmhamdi}, {Danziger}, {Branch}, \&
  {Leibundgut}}]{Elmhamdi07}
{Elmhamdi}, A., {Danziger}, I.~J., {Branch}, D., \& {Leibundgut}, B. 2007, in
  American Institute of Physics Conference Series, Vol. 924, The Multicolored
  Landscape of Compact Objects and Their Explosive Origins, ed. T.~{di Salvo},
  G.~L. {Israel}, L.~{Piersant}, L.~{Burderi}, G.~{Matt}, A.~{Tornambe}, \&
  M.~T. {Menna}, 277--284

\bibitem[{{Elmhamdi} {et~al.}(2006){Elmhamdi}, {Danziger}, {Branch},
  {Leibundgut}, {Baron}, \& {Kirshner}}]{Elmhamdi06}
{Elmhamdi}, A., {Danziger}, I.~J., {Branch}, D., {Leibundgut}, B., {Baron}, E.,
  \& {Kirshner}, R.~P. 2006, A\&A, 450, 305

\bibitem[{{Elmhamdi} {et~al.}(2004){Elmhamdi}, {Danziger}, {Cappellaro}, {Della
  Valle}, {Gouiffes}, {Phillips}, \& {Turatto}}]{Elmhamdi04}
{Elmhamdi}, A., {Danziger}, I.~J., {Cappellaro}, E., {Della Valle}, M.,
  {Gouiffes}, C., {Phillips}, M.~M., \& {Turatto}, M. 2004, \aap, 426, 963

\bibitem[{{Faran} {et~al.}(2014){Faran}, {Poznanski}, {Filippenko}, {Chornock},
  {Foley}, {Ganeshalingam}, {Leonard}, {Li}, {Modjaz}, {Serduke}, \&
  {Silverman}}]{Faran14}
{Faran}, T., {et~al.} 2014, MNRAS, 445, 554

\bibitem[{{Filippenko}(1988)}]{Filippenko88}
{Filippenko}, A.~V. 1988, Proceedings of the Astronomical Society of Australia,
  7, 540

\bibitem[{{Filippenko}(1997)}]{Filippenko97}
---. 1997, ARA\&A, 35, 309

\bibitem[{{Filippenko} {et~al.}(1990){Filippenko}, {Porter}, \&
  {Sargent}}]{Filippenko90}
{Filippenko}, A.~V., {Porter}, A.~C., \& {Sargent}, W.~L.~W. 1990, AJ, 100,
  1575

\bibitem[{{Filippenko} {et~al.}(1995){Filippenko}, {Barth}, {Matheson},
  {Armus}, {Brown}, {Espey}, {Fan}, {Goodrich}, {Ho}, {Junkkarinen}, {Koo},
  {Lehnert}, {Martel}, {Mazzarella}, {Miller}, {Smith}, {Tytler}, \&
  {Wirth}}]{Filippenko95}
{Filippenko}, A.~V., {et~al.} 1995, \apjl, 450, L11

\bibitem[{{Folatelli} {et~al.}(2014){Folatelli}, {Bersten}, {Kuncarayakti},
  {Olivares Estay}, {Anderson}, {Holmbo}, {Maeda}, {Morrell}, {Nomoto},
  {Pignata}, {Stritzinger}, {Contreras}, {F{\"o}rster}, {Hamuy}, {Phillips},
  {Prieto}, {Valenti}, {Afonso}, {Altenm{\"u}ller}, {Elliott}, {Greiner},
  {Updike}, {Haislip}, {LaCluyze}, {Moore}, \& {Reichart}}]{Folatelli14}
{Folatelli}, G., {et~al.} 2014, ApJ, 792, 7

\bibitem[{{Foley} {et~al.}(2003){Foley}, {Papenkova}, {Swift}, {Filippenko},
  {Li}, {Mazzali}, {Chornock}, {Leonard}, \& {Van Dyk}}]{Foley03}
{Foley}, R.~J., {et~al.} 2003, PASP, 115, 1220

\bibitem[{{Foley} {et~al.}(2009){Foley}, {Chornock}, {Filippenko},
  {Ganeshalingam}, {Kirshner}, {Li}, {Cenko}, {Challis}, {Friedman}, {Modjaz},
  {Silverman}, \& {Wood-Vasey}}]{Foley09}
---. 2009, AJ, 138, 376

\bibitem[{{Friesen} {et~al.}(2014){Friesen}, {Baron}, {Wisniewski}, {Parrent},
  {Thomas}, {Miller}, \& {Marion}}]{Friesen14}
{Friesen}, B., {Baron}, E., {Wisniewski}, J.~P., {Parrent}, J.~T., {Thomas},
  R.~C., {Miller}, T.~R., \& {Marion}, G.~H. 2014, ApJ, 792, 120

\bibitem[{{Fryer}(2004)}]{Fryer04}
{Fryer}, C.~L., ed. 2004, Astrophysics and Space Science Library, Vol. 302,
  {Stellar Collapse}

\bibitem[{{Gal-Yam} {et~al.}(2013){Gal-Yam}, {Mazzali}, {Manulis}, \&
  {Bishop}}]{Galyam13}
{Gal-Yam}, A., {Mazzali}, P.~A., {Manulis}, I., \& {Bishop}, D. 2013, PASP,
  125, 749

\bibitem[{{Gal-Yam} {et~al.}(2009){Gal-Yam}, {Mazzali}, {Ofek}, {Nugent},
  {Kulkarni}, {Kasliwal}, {Quimby}, {Filippenko}, {Cenko}, {Chornock},
  {Waldman}, {Kasen}, {Sullivan}, {Beshore}, {Drake}, {Thomas}, {Bloom},
  {Poznanski}, {Miller}, {Foley}, {Silverman}, {Arcavi}, {Ellis}, \&
  {Deng}}]{Galyam09}
{Gal-Yam}, A., {et~al.} 2009, Nature, 462, 624

\bibitem[{{Gall} {et~al.}(2015){Gall}, {Polshaw}, {Kotak}, {Jerkstrand},
  {Leibundgut}, {Rabinowitz}, {Sollerman}, {Sullivan}, {Smartt}, {Anderson},
  {Benetti}, {Baltay}, {Feindt}, {Fraser}, {Gonz{\'a}lez-Gait{\'a}n},
  {Inserra}, {Maguire}, {McKinnon}, {Valenti}, \& {Young}}]{Gall15}
{Gall}, E.~E.~E., {et~al.} 2015, arXiv:1502.06034

\bibitem[{{Ganeshalingam} {et~al.}(2012){Ganeshalingam}, {Li}, {Filippenko},
  {Silverman}, {Chornock}, {Foley}, {Matheson}, {Kirshner}, {Milne}, {Calkins},
  \& {Shen}}]{Ganeshalignam12}
{Ganeshalingam}, M., {et~al.} 2012, ApJ, 751, 142

\bibitem[{{Gaposchkin}(1936)}]{Payne36}
{Gaposchkin}, C.~P. 1936, ApJ, 83, 245

\bibitem[{{Garnavich} {et~al.}(2004){Garnavich}, {Bonanos}, {Krisciunas},
  {Jha}, {Kirshner}, {Schlegel}, {Challis}, {Macri}, {Hatano}, {Branch},
  {Bothun}, \& {Freedman}}]{Garnavich04}
{Garnavich}, P.~M., {et~al.} 2004, ApJ, 613, 1120

\bibitem[{{Gray} \& {Corbally}(2009)}]{Gray09}
{Gray}, R.~O., \& {Corbally}, J., C. 2009, {Stellar Spectral Classification}

\bibitem[{{Greiner} {et~al.}(2015){Greiner}, {Mazzali}, {Kann}, {Kr{\"u}hler},
  {Pian}, {Prentice}, {Olivares E.}, {Rossi}, {Klose}, {Taubenberger}, {Knust},
  {Afonso}, {Ashall}, {Bolmer}, {Delvaux}, {Diehl}, {Elliott}, {Filgas},
  {Fynbo}, {Graham}, {Guelbenzu}, {Kobayashi}, {Leloudas}, {Savaglio},
  {Schady}, {Schmidl}, {Schweyer}, {Sudilovsky}, {Tanga}, {Updike}, {van
  Eerten}, \& {Varela}}]{Greiner15}
{Greiner}, J., {et~al.} 2015, Nature, 523, 189

\bibitem[{{Hachinger} {et~al.}(2012){Hachinger}, {Mazzali}, {Taubenberger},
  {Fink}, {Pakmor}, {Hillebrandt}, \& {Seitenzahl}}]{Hachinger12}
{Hachinger}, S., {Mazzali}, P.~A., {Taubenberger}, S., {Fink}, M., {Pakmor},
  R., {Hillebrandt}, W., \& {Seitenzahl}, I.~R. 2012, MNRAS, 427, 2057

\bibitem[{{Hatano} {et~al.}(1999){Hatano}, {Branch}, {Fisher}, {Millard}, \&
  {Baron}}]{HatanoAtlas}
{Hatano}, K., {Branch}, D., {Fisher}, A., {Millard}, J., \& {Baron}, E. 1999,
  ApJS, 121, 233

\bibitem[{{Hatano} {et~al.}(2002){Hatano}, {Branch}, {Qiu}, {Baron},
  {Thielemann}, \& {Fisher}}]{Hatano02}
{Hatano}, K., {Branch}, D., {Qiu}, Y.~L., {Baron}, E., {Thielemann}, F., \&
  {Fisher}, A. 2002, Nature, 7, 441

\bibitem[{{Hillier} \& {Dessart}(2012)}]{Hillier12}
{Hillier}, D.~J., \& {Dessart}, L. 2012, MNRAS, 424, 252

\bibitem[{{Inserra} {et~al.}(2013){Inserra}, {Smartt}, {Jerkstrand}, {Valenti},
  {Fraser}, {Wright}, {Smith}, {Chen}, {Kotak}, {Pastorello}, {Nicholl},
  {Bresolin}, {Kudritzki}, {Benetti}, {Botticella}, {Burgett}, {Chambers},
  {Ergon}, {Flewelling}, {Fynbo}, {Geier}, {Hodapp}, {Howell}, {Huber},
  {Kaiser}, {Leloudas}, {Magill}, {Magnier}, {McCrum}, {Metcalfe}, {Price},
  {Rest}, {Sollerman}, {Sweeney}, {Taddia}, {Taubenberger}, {Tonry},
  {Wainscoat}, {Waters}, \& {Young}}]{Inserra13}
{Inserra}, C., {et~al.} 2013, ApJ, 770, 128

\bibitem[{{Iwamoto} {et~al.}(1994){Iwamoto}, {Nomoto}, {H{\"o}flich},
  {Yamaoka}, {Kumagai}, \& {Shigeyama}}]{Iwamoto94}
{Iwamoto}, K., {Nomoto}, K., {H{\"o}flich}, P., {Yamaoka}, H., {Kumagai}, S.,
  \& {Shigeyama}, T. 1994, \apjl, 437, L115

\bibitem[{{Iwamoto} {et~al.}(1998){Iwamoto}, {Mazzali}, {Nomoto}, {Umeda},
  {Nakamura}, {Patat}, {Danziger}, {Young}, {Suzuki}, {Shigeyama},
  {Augusteijn}, {Doublier}, {Gonzalez}, {Boehnhardt}, {Brewer}, {Hainaut},
  {Lidman}, {Leibundgut}, {Cappellaro}, {Turatto}, {Galama}, {Vreeswijk},
  {Kouveliotou}, {van Paradijs}, {Pian}, {Palazzi}, \& {Frontera}}]{Iwamoto98}
{Iwamoto}, K., {et~al.} 1998, Nature, 395, 672

\bibitem[{{James} \& {Baron}(2010)}]{James10}
{James}, S., \& {Baron}, E. 2010, ApJ, 718, 957

\bibitem[{{Jeffery} {et~al.}(1991){Jeffery}, {Branch}, {Filippenko}, \&
  {Nomoto}}]{Jeffery87M91}
{Jeffery}, D.~J., {Branch}, D., {Filippenko}, A.~V., \& {Nomoto}, K. 1991,
  \apjl, 377, L89

\bibitem[{{Jeffery} {et~al.}(2007){Jeffery}, {Ketchum}, {Branch}, {Baron},
  {Elmhamdi}, \& {Danziger}}]{Jeffery07}
{Jeffery}, D.~J., {Ketchum}, W., {Branch}, D., {Baron}, E., {Elmhamdi}, A., \&
  {Danziger}, I.~J. 2007, ApJS, 171, 493

\bibitem[{{Jerkstrand} {et~al.}(2015){Jerkstrand}, {Ergon}, {Smartt},
  {Fransson}, {Sollerman}, {Taubenberger}, {Bersten}, \&
  {Spyromilio}}]{Jerkstrand15}
{Jerkstrand}, A., {Ergon}, M., {Smartt}, S.~J., {Fransson}, C., {Sollerman},
  J., {Taubenberger}, S., {Bersten}, M., \& {Spyromilio}, J. 2015, aap, 573,
  A12

\bibitem[{{Kerzendorf} \& {Sim}(2014)}]{Kerzendorf14}
{Kerzendorf}, W.~E., \& {Sim}, S.~A. 2014, MNRAS, 440, 387

\bibitem[{{Ketchum} {et~al.}(2008){Ketchum}, {Baron}, \& {Branch}}]{Ketchum08}
{Ketchum}, W., {Baron}, E., \& {Branch}, D. 2008, ApJ, 674, 371

\bibitem[{{Kromer} {et~al.}(2013){Kromer}, {Fink}, {Stanishev}, {Taubenberger},
  {Ciaraldi-Schoolman}, {Pakmor}, {R{\"o}pke}, {Ruiter}, {Seitenzahl}, {Sim},
  {Blanc}, {Elias-Rosa}, \& {Hillebrandt}}]{Kromer13}
{Kromer}, M., {et~al.} 2013, MNRAS, 429, 2287

\bibitem[{{Leloudas} {et~al.}(2015){Leloudas}, {Patat}, {Maund}, {Hsiao},
  {Malesani}, {Schulze}, {Contreras}, {de Ugarte Postigo}, {Sollerman},
  {Stritzinger}, {Taddia}, {Wheeler}, \& {Gorosabel}}]{Leloudas15LSQ14mo}
{Leloudas}, G., {et~al.} 2015, ApJL, 815, L10

\bibitem[{{Lentz} {et~al.}(2001){Lentz}, {Baron}, {Branch}, \&
  {Hauschildt}}]{Lentz01a}
{Lentz}, E.~J., {Baron}, E., {Branch}, D., \& {Hauschildt}, P.~H. 2001, ApJ,
  557, 266

\bibitem[{{Li} {et~al.}(2003){Li}, {Filippenko}, {Chornock}, {Berger},
  {Berlind}, {Calkins}, {Challis}, {Fassnacht}, {Jha}, {Kirshner}, {Matheson},
  {Sargent}, {Simcoe}, {Smith}, \& {Squires}}]{WLi03}
{Li}, W., {et~al.} 2003, PASP, 115, 453

\bibitem[{{Liu} {et~al.}(2015){Liu}, {Modjaz}, {Bianco}, \& {Graur}}]{YLiu15}
{Liu}, Y.-Q., {Modjaz}, M., {Bianco}, F.~B., \& {Graur}, O. 2015,
  arXiv:1510.08049

\bibitem[{{Lyman} {et~al.}(2014){Lyman}, {Bersier}, {James}, {Mazzali},
  {Eldridge}, {Fraser}, \& {Pian}}]{Lyman14}
{Lyman}, J., {Bersier}, D., {James}, P., {Mazzali}, P., {Eldridge}, J.,
  {Fraser}, M., \& {Pian}, E. 2014, arXiv:1406.3667

\bibitem[{{Maguire} {et~al.}(2010){Maguire}, {Di Carlo}, {Smartt},
  {Pastorello}, {Tsvetkov}, {Benetti}, {Spiro}, {Arkharov}, {Beccari},
  {Botticella}, {Cappellaro}, {Cristallo}, {Dolci}, {Elias-Rosa}, {Fiaschi},
  {Gorshanov}, {Harutyunyan}, {Larionov}, {Navasardyan}, {Pietrinferni},
  {Raimondo}, {di Rico}, {Valenti}, {Valentini}, \& {Zampieri}}]{Maguire10}
{Maguire}, K., {et~al.} 2010, \mnras, 404, 981

\bibitem[{{Marion} {et~al.}(2013){Marion}, {Vinko}, {Wheeler}, {Foley},
  {Hsiao}, {Brown}, {Challis}, {Filippenko}, {Garnavich}, {Kirshner},
  {Landsman}, {Parrent}, {Pritchard}, {Roming}, {Silverman}, \&
  {Wang}}]{Marion13}
{Marion}, G.~H., {et~al.} 2013, ApJ, 777, 40

\bibitem[{{Matheson} {et~al.}(2001){Matheson}, {Filippenko}, {Li}, {Leonard},
  \& {Shields}}]{Matheson01}
{Matheson}, T., {Filippenko}, A.~V., {Li}, W., {Leonard}, D.~C., \& {Shields},
  J.~C. 2001, \aj, 121, 1648

\bibitem[{{Maund} {et~al.}(2013){Maund}, {Spyromilio}, {H{\"o}flich},
  {Wheeler}, {Baade}, {Clocchiatti}, {Patat}, {Reilly}, {Wang}, \&
  {Zelaya}}]{Maund13}
{Maund}, J.~R., {et~al.} 2013, MNRAS, 433, L20

\bibitem[{{Mazzali} {et~al.}(2000){Mazzali}, {Iwamoto}, \&
  {Nomoto}}]{Mazzali97ef00}
{Mazzali}, P.~A., {Iwamoto}, K., \& {Nomoto}, K. 2000, ApJ, 545, 407

\bibitem[{{Mazzali} {et~al.}(2013){Mazzali}, {Walker}, {Pian}, {Tanaka},
  {Corsi}, {Hattori}, \& {Gal-Yam}}]{Mazzali13}
{Mazzali}, P.~A., {Walker}, E.~S., {Pian}, E., {Tanaka}, M., {Corsi}, A.,
  {Hattori}, T., \& {Gal-Yam}, A. 2013, \mnras, 432, 2463

\bibitem[{{Mazzali} {et~al.}(2003){Mazzali}, {Deng}, {Tominaga}, {Maeda},
  {Nomoto}, {Matheson}, {Kawabata}, {Stanek}, \& {Garnavich}}]{Mazzali03}
{Mazzali}, P.~A., {et~al.} 2003, \apjl, 599, L95

\bibitem[{{Mazzali} {et~al.}(2006){Mazzali}, {Deng}, {Pian}, {Malesani},
  {Tominaga}, {Maeda}, {Nomoto}, {Chincarini}, {Covino}, {Della Valle},
  {Fugazza}, {Tagliaferri}, \& {Gal-Yam}}]{Mazzali06}
---. 2006, \apj, 645, 1323

\bibitem[{{Mazzali} {et~al.}(2014){Mazzali}, {Sullivan}, {Hachinger}, {Ellis},
  {Nugent}, {Howell}, {Gal-Yam}, {Maguire}, {Cooke}, {Thomas}, {Nomoto}, \&
  {Walker}}]{Mazzali14}
---. 2014, MNRAS, 439, 1959

\bibitem[{{Milisavljevic} {et~al.}(2013){Milisavljevic}, {Margutti},
  {Soderberg}, {Pignata}, {Chomiuk}, {Fesen}, {Bufano}, {Sanders}, {Parrent},
  {Parker}, {Mazzali}, {Pian}, {Pickering}, {Buckley}, {Crawford}, {Gulbis},
  {Hettlage}, {Hooper}, {Nordsieck}, {O'Donoghue}, {Husser}, {Potter},
  {Kniazev}, {Kotze}, {Romero-Colmenero}, {Vaisanen}, {Wolf}, {Bietenholz},
  {Bartel}, {Fransson}, {Walker}, {Brunthaler}, {Chakraborti}, {Levesque},
  {MacFadyen}, {Drescher}, {Bock}, {Marples}, {Anderson}, {Benetti},
  {Reichart}, \& {Ivarsen}}]{Danmil1311ei}
{Milisavljevic}, D., {et~al.} 2013, ApJ, 767, 71

\bibitem[{{Milisavljevic} {et~al.}(2015){Milisavljevic}, {Margutti}, {Kamble},
  {Patnaude}, {Raymond}, {Eldridge}, {Fong}, {Bietenholz}, {Challis},
  {Chornock}, {Drout}, {Fransson}, {Fesen}, {Grindlay}, {Kirshner}, {Lunnan},
  {Mackey}, {Miller}, {Parrent}, {Sanders}, {Soderberg}, \&
  {Zauderer}}]{danmil15}
---. 2015, ApJ, 815, 120

\bibitem[{{Minkowski}(1941)}]{Minkowski41}
{Minkowski}, R. 1941, PASP, 53, 224

\bibitem[{{Modjaz} {et~al.}(2009){Modjaz}, {Li}, {Butler}, {Chornock},
  {Perley}, {Blondin}, {Bloom}, {Filippenko}, {Kirshner}, {Kocevski},
  {Poznanski}, {Hicken}, {Foley}, {Stringfellow}, {Berlind}, {Barrado y
  Navascues}, {Blake}, {Bouy}, {Brown}, {Challis}, {Chen}, {de Vries},
  {Dufour}, {Falco}, {Friedman}, {Ganeshalingam}, {Garnavich}, {Holden},
  {Illingworth}, {Lee}, {Liebert}, {Marion}, {Olivier}, {Prochaska},
  {Silverman}, {Smith}, {Starr}, {Steele}, {Stockton}, {Williams}, \&
  {Wood-Vasey}}]{Modjaz09}
{Modjaz}, M., {et~al.} 2009, ApJ, 702, 226

\bibitem[{{Modjaz} {et~al.}(2014){Modjaz}, {Blondin}, {Kirshner}, {Matheson},
  {Berlind}, {Bianco}, {Calkins}, {Challis}, {Garnavich}, {Hicken}, {Jha},
  {Liu}, \& {Marion}}]{Modjaz14}
---. 2014, AJ, 147, 99

\bibitem[{{Moriya} {et~al.}(2010){Moriya}, {Tominaga}, {Tanaka}, {Maeda}, \&
  {Nomoto}}]{Moriya10}
{Moriya}, T., {Tominaga}, N., {Tanaka}, M., {Maeda}, K., \& {Nomoto}, K. 2010,
  ApJL, 717, L83

\bibitem[{{Moriya} {et~al.}(2015){Moriya}, {Liu}, {Mackey}, {Chen}, \&
  {Langer}}]{Moriya15}
{Moriya}, T.~J., {Liu}, Z.-W., {Mackey}, J., {Chen}, T.-W., \& {Langer}, N.
  2015, A\&A, 584, L5

\bibitem[{{Nicholl} {et~al.}(2014){Nicholl}, {Smartt}, {Jerkstrand}, {Inserra},
  {Anderson}, {Baltay}, {Benetti}, {Chen}, {Elias-Rosa}, {Feindt}, {Fraser},
  {Gal-Yam}, {Hadjiyska}, {Howell}, {Kotak}, {Lawrence}, {Leloudas},
  {Margheim}, {Mattila}, {McCrum}, {McKinnon}, {Mead}, {Nugent}, {Rabinowitz},
  {Rest}, {Smith}, {Sollerman}, {Sullivan}, {Taddia}, {Valenti}, {Walker}, \&
  {Young}}]{Nicholl14}
{Nicholl}, M., {et~al.} 2014, \mnras, 444, 2096

\bibitem[{{Nicholl} {et~al.}(2015){Nicholl}, {Smartt}, {Jerkstrand}, {Inserra},
  {Sim}, {Chen}, {Benetti}, {Fraser}, {Gal-Yam}, {Kankare}, {Maguire}, {Smith},
  {Sullivan}, {Valenti}, {Young}, {Baltay}, {Bauer}, {Baumont}, {Bersier},
  {Botticella}, {Childress}, {Dennefeld}, {Della Valle}, {Elias-Rosa},
  {Feindt}, {Galbany}, {Hadjiyska}, {Le Guillou}, {Leloudas}, {Mazzali},
  {McKinnon}, {Polshaw}, {Rabinowitz}, {Rostami}, {Scalzo}, {Schmidt},
  {Schulze}, {Sollerman}, {Taddia}, \& {Yuan}}]{Nicholl15}
---. 2015, MNRAS, 452, 3869

\bibitem[{{Nomoto} {et~al.}(2000){Nomoto}, {Maeda}, {Nakamura}, {Iwamoto},
  {Suzuki}, {Mazzali}, {Turatto}, {Danziger}, \& {Patat}}]{Nomoto00}
{Nomoto}, K., {et~al.} 2000, in American Institute of Physics Conference
  Series, Vol. 526, Gamma-ray Bursts, 5th Huntsville Symposium, ed. R.~M.
  {Kippen}, R.~S. {Mallozzi}, \& G.~J. {Fishman}, 622--627

\bibitem[{{Parrent} {et~al.}(2014){Parrent}, {Friesen}, \&
  {Parthasarathy}}]{Parrent14}
{Parrent}, J., {Friesen}, B., \& {Parthasarathy}, M. 2014, Ap\&SS, 351, 1

\bibitem[{{Parrent} {et~al.}(2007){Parrent}, {Branch}, {Troxel}, {Casebeer},
  {Jeffery}, {Ketchum}, {Baron}, {Serduke}, \& {Filippenko}}]{Parrent07}
{Parrent}, J., {et~al.} 2007, PASP, 119, 135

\bibitem[{{Parrent} {et~al.}(2011){Parrent}, {Thomas}, {Fesen}, {Marion},
  {Challis}, {Garnavich}, {Milisavljevic}, {Vink{\`o}}, \&
  {Wheeler}}]{Parrent11}
{Parrent}, J.~T., {et~al.} 2011, ApJ, 732, 30

\bibitem[{{Parrent} {et~al.}(2012){Parrent}, {Howell}, {Friesen}, {Thomas},
  {Fesen}, {Milisavljevic}, {Bianco}, {Dilday}, {Nugent}, {Baron}, {Arcavi},
  {Ben-Ami}, {Bersier}, {Bildsten}, {Bloom}, {Cao}, {Cenko}, {Filippenko},
  {Gal-Yam}, {Kasliwal}, {Konidaris}, {Kulkarni}, {Law}, {Levitan}, {Maguire},
  {Mazzali}, {Ofek}, {Pan}, {Polishook}, {Poznanski}, {Quimby}, {Silverman},
  {Sternberg}, {Sullivan}, {Walker}, {Xu}, {Buton}, \& {Pereira}}]{Parrent12}
---. 2012, ApJL, 752, L26

\bibitem[{{Patat} {et~al.}(1996){Patat}, {Benetti}, {Cappellaro}, {Danziger},
  {della Valle}, {Mazzali}, \& {Turatto}}]{Patat96}
{Patat}, F., {Benetti}, S., {Cappellaro}, E., {Danziger}, I.~J., {della Valle},
  M., {Mazzali}, P.~A., \& {Turatto}, M. 1996, MNRAS, 278, 111

\bibitem[{{Patat} {et~al.}(2001){Patat}, {Cappellaro}, {Danziger}, {Mazzali},
  {Sollerman}, {Augusteijn}, {Brewer}, {Doublier}, {Gonzalez}, {Hainaut},
  {Lidman}, {Leibundgut}, {Nomoto}, {Nakamura}, {Spyromilio}, {Rizzi},
  {Turatto}, {Walsh}, {Galama}, {van Paradijs}, {Kouveliotou}, {Vreeswijk},
  {Frontera}, {Masetti}, {Palazzi}, \& {Pian}}]{Patat01}
{Patat}, F., {et~al.} 2001, ApJ, 555, 900

\bibitem[{{Pejcha} \& {Prieto}(2015)}]{Pejcha15}
{Pejcha}, O., \& {Prieto}, J.~L. 2015, arXiv:1501.06573

\bibitem[{{Pereira} {et~al.}(2013){Pereira}, {Thomas}, {Aldering}, {Antilogus},
  {Baltay}, {Benitez-Herrera}, {Bongard}, {Buton}, {Canto}, {Cellier-Holzem},
  {Chen}, {Childress}, {Chotard}, {Copin}, {Fakhouri}, {Fink}, {Fouchez},
  {Gangler}, {Guy}, {Hillebrandt}, {Hsiao}, {Kerschhaggl}, {Kowalski},
  {Kromer}, {Nordin}, {Nugent}, {Paech}, {Pain}, {P{\'e}contal}, {Perlmutter},
  {Rabinowitz}, {Rigault}, {Runge}, {Saunders}, {Smadja}, {Tao},
  {Taubenberger}, {Tilquin}, \& {Wu}}]{Pereira13}
{Pereira}, R., {et~al.} 2013, A\&A, 554, A27

\bibitem[{{Popper}(1937)}]{Popper37}
{Popper}, D.~M. 1937, PASP, 49, 283

\bibitem[{{Pun} {et~al.}(1995){Pun}, {Kirshner}, {Sonneborn}, {Challis},
  {Nassiopoulos}, {Arquilla}, {Crenshaw}, {Shrader}, {Teays}, {Cassatella},
  {Gilmozzi}, {Talavera}, {Wamsteker}, {Fransson}, \& {Panagia}}]{Pun95}
{Pun}, C.~S.~J., {et~al.} 1995, \apjs, 99, 223

\bibitem[{{Richardson} {et~al.}(2001){Richardson}, {Thomas}, {Casebeer},
  {Blankenship}, {Ratowt}, {Baron}, \& {Branch}}]{Richardson01}
{Richardson}, D., {Thomas}, R.~C., {Casebeer}, D., {Blankenship}, Z., {Ratowt},
  S., {Baron}, E., \& {Branch}, D. 2001, in Bulletin of the American
  Astronomical Society, Vol.~33, American Astronomical Society Meeting
  Abstracts, 1428

\bibitem[{{Sanders} {et~al.}(2014){Sanders}, {Soderberg}, {Gezari},
  {Betancourt}, {Chornock}, {Berger}, {Foley}, {Challis}, {Drout}, {Kirshner},
  {Lunnan}, {Marion}, {Margutti}, {McKinnon}, {Milisavljevic}, {Narayan},
  {Rest}, {Kankare}, {Mattila}, {Smartt}, {Huber}, {Burgett}, {Draper},
  {Hodapp}, {Kaiser}, {Kudritzki}, {Magnier}, {Metcalfe}, {Morgan}, {Price},
  {Tonry}, {Wainscoat}, \& {Waters}}]{Sanders14}
{Sanders}, N.~E., {et~al.} 2014, arXiv:1404.2004

\bibitem[{{Sasdelli} {et~al.}(2014){Sasdelli}, {Mazzali}, {Pian}, {Nomoto},
  {Hachinger}, {Cappellaro}, \& {Benetti}}]{Sasdelli14}
{Sasdelli}, M., {Mazzali}, P.~A., {Pian}, E., {Nomoto}, K., {Hachinger}, S.,
  {Cappellaro}, E., \& {Benetti}, S. 2014, MNRAS, 445, 711

\bibitem[{{Sauer} {et~al.}(2006){Sauer}, {Hoffmann}, \& {Pauldrach}}]{Sauer06}
{Sauer}, D.~N., {Hoffmann}, T.~L., \& {Pauldrach}, A.~W.~A. 2006, A\&A, 459,
  229

\bibitem[{{Smith} {et~al.}(2007){Smith}, {Li}, {Foley}, {Wheeler}, {Pooley},
  {Chornock}, {Filippenko}, {Silverman}, {Quimby}, {Bloom}, \&
  {Hansen}}]{Smith07}
{Smith}, N., {et~al.} 2007, \apj, 666, 1116

\bibitem[{{Soderberg} {et~al.}(2008){Soderberg}, {Berger}, {Page}, {Schady},
  {Parrent}, {Pooley}, {Wang}, {Ofek}, {Cucchiara}, {Rau}, {Waxman}, {Simon},
  {Bock}, {Milne}, {Page}, {Barentine}, {Barthelmy}, {Beardmore}, {Bietenholz},
  {Brown}, {Burrows}, {Burrows}, {Byrngelson}, {Cenko}, {Chandra}, {Cummings},
  {Fox}, {Gal-Yam}, {Gehrels}, {Immler}, {Kasliwal}, {Kong}, {Krimm},
  {Kulkarni}, {Maccarone}, {M{\'e}sz{\'a}ros}, {Nakar}, {O'Brien}, {Overzier},
  {de Pasquale}, {Racusin}, {Rea}, \& {York}}]{Soderberg08D}
{Soderberg}, A.~M., {et~al.} 2008, Nature, 453, 469

\bibitem[{{Stathakis} {et~al.}(2000){Stathakis}, {Boyle}, {Jones}, {Bessell},
  {Galama}, {Germany}, {Hartley}, {James}, {Kouveliotou}, {Lewis}, {Parker},
  {Russell}, {Sadler}, {Tinney}, {van Paradijs}, \& {Vreeswijk}}]{Stathakis00}
{Stathakis}, R.~A., {et~al.} 2000, MNRAS, 314, 807

\bibitem[{{Stritzinger} {et~al.}(2014){Stritzinger}, {Hsiao}, {Valenti},
  {Taddia}, {Rivera-Thorsen}, {Leloudas}, {Maeda}, {Pastorello}, {Phillips},
  {Pignata}, {Baron}, {Burns}, {Contreras}, {Folatelli}, {Hamuy},
  {H{\"o}flich}, {Morrell}, {Prieto}, {Benetti}, {Campillay}, {Haislip},
  {LaClutze}, {Moore}, \& {Reichart}}]{Stritzinger14}
{Stritzinger}, M.~D., {et~al.} 2014, A\&A, 561, A146

\bibitem[{{Taubenberger} {et~al.}(2006){Taubenberger}, {Pastorello}, {Mazzali},
  {Valenti}, {Pignata}, {Sauer}, {Arbey}, {B{\"a}rnbantner}, {Benetti}, {Della
  Valle}, {Deng}, {Elias-Rosa}, {Filippenko}, {Foley}, {Goobar}, {Kotak}, {Li},
  {Meikle}, {Mendez}, {Patat}, {Pian}, {Ries}, {Ruiz-Lapuente}, {Salvo},
  {Stanishev}, {Turatto}, \& {Hillebrandt}}]{Taubenberger06}
{Taubenberger}, S., {et~al.} 2006, \mnras, 371, 1459

\bibitem[{{Taubenberger} {et~al.}(2011){Taubenberger}, {Benetti}, {Childress},
  {Pakmor}, {Hachinger}, {Mazzali}, {Stanishev}, {Elias-Rosa}, {Agnoletto},
  {Bufano}, {Ergon}, {Harutyunyan}, {Inserra}, {Kankare}, {Kromer},
  {Navasardyan}, {Nicolas}, {Pastorello}, {Prosperi}, {Salgado}, {Sollerman},
  {Stritzinger}, {Turatto}, {Valenti}, \& {Hillebrandt}}]{Taubenberger11}
---. 2011, MNRAS, 412, 2735

\bibitem[{{Toy} {et~al.}(2015){Toy}, {Cenko}, {Silverman}, {Butler},
  {Cucchiara}, {Watson}, {Bersier}, {Perley}, {Margutti}, {Bellm}, {Bloom},
  {Cao}, {Capone}, {Clubb}, {Corsi}, {de Diego}, {Filippenko}, {Fox},
  {Gal-Yam}, {Gehrels}, {Georgiev}, {Gonz{\'a}lez}, {Kasliwal}, {Kelly},
  {Kulkarni}, {Kutyrev}, {Lee}, {Prochaska}, {Ramirez-Ruiz}, {Richer},
  {Rom{\'a}n}, {Singer}, {Stern}, {Troja}, \& {Veilleux}}]{Toy15}
{Toy}, V.~L., {et~al.} 2015, arXiv:1508.00575

\bibitem[{{Tsvetkov}(1987)}]{Tsvetkov87}
{Tsvetkov}, D.~Y. 1987, Soviet Astronomy Letters, 13, 376

\bibitem[{{Vacca} {et~al.}(2015){Vacca}, {Hamilton}, {Savage}, {Shenoy},
  {Becklin}, {McLean}, {Logsdon}, {Marion}, {Ashok}, {Banerjee}, {Evans},
  {Fox}, {Garnavich}, {Gehrz}, {Greenhouse}, {Helton}, {Kirshner}, {Shenoy},
  {Smith}, {Spyromilio}, {Starrfield}, {Wooden}, \& {Woodward}}]{Vacca15}
{Vacca}, W.~D., {et~al.} 2015, \apj, 804, 66

\bibitem[{{Valenti} {et~al.}(2011){Valenti}, {Fraser}, {Benetti}, {Pignata},
  {Sollerman}, {Inserra}, {Cappellaro}, {Pastorello}, {Smartt}, {Ergon},
  {Botticella}, {Brimacombe}, {Bufano}, {Crockett}, {Eder}, {Fugazza},
  {Haislip}, {Hamuy}, {Harutyunyan}, {Ivarsen}, {Kankare}, {Kotak}, {Lacluyze},
  {Magill}, {Mattila}, {Maza}, {Mazzali}, {Reichart}, {Taubenberger},
  {Turatto}, \& {Zampieri}}]{Valenti11}
{Valenti}, S., {et~al.} 2011, MNRAS, 416, 3138

\bibitem[{{Wheeler} {et~al.}(1994){Wheeler}, {Harkness}, {Clocchiatti},
  {Benetti}, {Brotherton}, {Depoy}, \& {Elias}}]{Wheeler94}
{Wheeler}, J.~C., {Harkness}, R.~P., {Clocchiatti}, A., {Benetti}, S.,
  {Brotherton}, M.~S., {Depoy}, D.~L., \& {Elias}, J. 1994, \apjl, 436, L135

\bibitem[{{Wheeler} {et~al.}(1995){Wheeler}, {Harkness}, {Khokhlov}, \&
  {Hoeflich}}]{Wheeler95}
{Wheeler}, J.~C., {Harkness}, R.~P., {Khokhlov}, A.~M., \& {Hoeflich}, P. 1995,
  Physics Reports, 256, 211

\bibitem[{{Wheeler} {et~al.}(2015){Wheeler}, {Johnson}, \&
  {Clocchiatti}}]{Wheeler15}
{Wheeler}, J.~C., {Johnson}, V., \& {Clocchiatti}, A. 2015, \mnras, 450, 1295

\bibitem[{{Wheeler} \& {Levreault}(1985)}]{Wheeler85}
{Wheeler}, J.~C., \& {Levreault}, R. 1985, ApJL, 294, L17

\bibitem[{{White} {et~al.}(2015){White}, {Kasliwal}, {Nugent}, {Gal-Yam},
  {Howell}, {Sullivan}, {Goobar}, {Piro}, {Bloom}, {Kulkarni}, {Laher},
  {Masci}, {Ofek}, {Surace}, {Ben-Ami}, {Cao}, {Cenko}, {Hook}, {J{\"o}nsson},
  {Matheson}, {Sternberg}, {Quimby}, \& {Yaron}}]{White15}
{White}, C.~J., {et~al.} 2015, ApJ, 799, 52

\bibitem[{{Yan} {et~al.}(2015){Yan}, {Quimby}, {Ofek}, {Gal-Yam}, {Mazzali},
  {Perley}, {Vreeswijk}, {Leloudas}, {de Cia}, {Masci}, {Cenko}, {Cao},
  {Kulkarni}, {Nugent}, {Rebbapragada}, {Wo{\'z}niak}, \& {Yaron}}]{Yan15}
{Yan}, L., {et~al.} 2015, ApJ, 814, 108

\bibitem[{{Yaron} \& {Gal-Yam}(2012)}]{WISEREP}
{Yaron}, O., \& {Gal-Yam}, A. 2012, PASP, 124, 668

\bibitem[{{Yoshida} \& {Umeda}(2011)}]{Yoshida11}
{Yoshida}, T., \& {Umeda}, H. 2011, \mnras, 412, L78

\end{thebibliography}


\begin{table*}
\centering
\caption{Common Lines in Optical Spectra of Type I Supernovae}
\begin{threeparttable}[b]
\begin{tabular}{lr}
 \tableline\tableline
Ion & Rest Wavelength (\AA) \\ 
\tableline
$^{a}$\ion{C}{2}$^{bc}$ & 6580, 7234 \\
$^{a}$\ion{O}{1}$^{bc}$ &7774, 8446, 9264 \\ 
$^{a}$\ion{Mg}{2}$^{bc}$ & 4481, 7896 \\
$^{a}$\ion{Ca}{2}$^{bc}$ & 3969, 3934, 8498, 8542, 8662 \\
$^{a}$\ion{Si}{2} & 3838, 4130, 5051, 5972, 6355* \\
$^{a}$\ion{Si}{3} & 4560, 5743 \\
$^{a}$\ion{S}{2} & 4163, 5032, 5208, 5468, 5654, 6305 \\
$^{a}$\ion{Fe}{2}$^{bc}$ & 4025, 4549, 4924, 5018, 5169 \\
$^{a}$\ion{Fe}{3} & 4420, 5156, 6000 \\
\tableline
\end{tabular}
\begin{tablenotes}
\item [abc] If the `a' is on the left, an ion has been detected for SN~Ia; ``bc'' on the right, an ion has been detected for SN~Ib~and~Ic. \ion{Fe}{3} may or may not be detectable, if present, on account of severe line-blending for all SN~Ib~and~Ic. For fast-evolving events, such as SN~2005hk and 2010X, an identification of \ion{Si}{2}~$\lambda$6355 also seems likely (Drout~et~al.~2013). \\
\item [*] When \ion{Si}{2} is identified in the spectra of SN~Ib, Ic, BL-Ic, and SLSN, it is based on a single line, $\lambda$6355, i.e. unconfirmed in the same sense as for H$\alpha$.
\end{tablenotes}
\end{threeparttable}
\end{table*}

\end{document}